\documentclass[usenatbib]{mn2e} 

\usepackage{caption}
\usepackage{tabularx}
\usepackage{lmodern}
\usepackage{color}
\usepackage{graphicx}
\usepackage{txfonts}
\usepackage{multirow}
\usepackage{supertabular}
\usepackage{longtable}
\usepackage{rotating}
\usepackage{rotating} 
\usepackage{lscape} 
\usepackage{amssymb}
\usepackage{dcolumn}

\usepackage{arydshln}
\usepackage{natbib}

\DeclareMathAlphabet{\mathsc}{OT1}{cmr}{m}{sc}
\def\testbx{bx}%
\DeclareRobustCommand{\ion}[2]{%
\relax\ifmmode
\ifx\testbx\f@series
{\mathbf{#1\,\mathsc{#2}}}\else
{\mathrm{#1\,\mathsc{#2}}}\fi
\else\textup{#1\,{\mdseries\textsc{#2}}}%
\fi}

\newcommand{\MgII}{\ion{Mg}{ii}}

\def\h2{$\rm H_2$}
\def\chin{$\chi^2_{\nu}$}
\def\sys{HE~0027$-$1836}
\def\lya{\ensuremath{{\rm Ly}\alpha}}

\def\lyb{\ensuremath{{\rm Ly}\beta}}
\def\kms{km\,s$^{-1}$}
\def\ms{m\,s$^{-1}$}

\def\zabs{$z_{\rm abs}$}

\def\dmm{$\Delta \mu/\mu$}

\def\nhi{$N$(H~{\sc i})}
\begin{document}

\title[Constraining  \dmm\ towards \sys]
{The UVES Large Program for Testing Fundamental Physics II: \\
Constraints on a Change in $\mu$ Towards Quasar \sys
\thanks{Based on data obtained with 
UVES at the Very Large Telescope of the European Southern Observatory 
(Prgm. ID 185.A-0745)}}
\author[Rahmani et. al.]{H. Rahmani$^{1}$, M. Wendt$^{2,3}$, R. Srianand$^{1}$, P. Noterdaeme$^{4}$, P. Petitjean$^{4}$, P. Molaro $^{5,6}$,  
\newauthor J. B. Whitmore$^{7}$, M. T. Murphy$^{7}$, M. Centurion$^{5}$, H. Fathivavsari$^{8}$, S. D'Odorico$^{9}$,  
\newauthor T. M. Evans$^{7}$, S. A. Levshakov$^{10,11}$, S. Lopez$^{12}$, C. J. A. P. Martins$^{6}$,  D. Reimers$^{2}$,  
\newauthor ~~~~~~~~~~~~~~~~~~~~~~~~~~~~~~~~~~~~~~~~~~~~ and G. Vladilo$^5$\\
$^{1}$ Inter-University Centre for Astronomy and Astrophysics, Post Bag 4,  Ganeshkhind, Pune 411\,007, India \\
$^{2}$ Hamburger Sternwarte, Universit\"{a}t Hamburg, Gojenbergsweg 112, 21029 Hamburg, Germany \\
$^{3}$ Institut f\"{u}r Physik und Astronomie, Universit\"{a}t Potsdam, 14476 Golm, Germany \\
$^{4}$ Institut d'Astrophysique de Paris, CNRS-UMPC, UMR7095, 98bis Bd Arago, 75014 Paris, France \\
$^{5}$ INAF-Osservatorio Astronomico di Trieste, Via G. B. Tiepolo 11, 34131 Trieste, Italy \\
$^{6}$ Centro de Astrof\'{i}sica, Universidade do Porto, Rua das Estrelas, 4150-762 Porto, Portugal \\
$^{7}$ Centre for Astrophysics and Supercomputing, Swinburne University of Technology, Hawthorn, VIC 3122, Australia\\
$^{8}$ Department of Theoretical Physics and Astrophysics, University of Tabriz P.O. Box 51664, Tabriz, Iran\\
$^{9}$ ESO, Karl Schwarzschild-Str. 1 85748 Garching, Germany\\
$^{10}$ Ioffe Physical-Technical Institute, Polytekhnicheskaya, Str. 26, 194021 Saint Petersburg, Russia\\
$^{11}$ St. Petersburg Electrotechnical University 'LETI', Prof. Popov Str. 5, 197376 St. Petersburg, Russia\\
$^{12}$ Departamento de Astronomia, Universidad de Chile, Casilla 36-D, Santiago, Chile\\
}
\pagerange{\pageref{firstpage}--\pageref{lastpage}} \pubyear{2012}
\maketitle
\label{firstpage}

\begin {abstract}
{
We present an accurate  analysis of the \h2\ absorption lines 
from the  \zabs\ $\sim$ 2.4018  damped \lya system towards \sys\ observed with the 
Very Large Telescope Ultraviolet and Visual Echelle Spectrograph (VLT/UVES) 
as a part of the European Southern Observatory Large Programme ''The UVES large programme for testing 
fundamental physics''  to constrain the variation of proton-to-electron mass ratio, $\mu \equiv m_p/m_e$. We perform cross-correlation analysis between
19 individual exposures taken over three years  and the combined
spectrum to check the wavelength calibration stability. 
We notice the presence of a 
possible wavelength dependent velocity drift especially in the 
data taken in 2012. 
We use available asteroids spectra taken with UVES close to our observations to 
confirm and quantify this effect. 
We consider single and two component Voigt profiles to model the observed 
\h2\ absorption profiles. We use both linear regression analysis and Voigt profile 
fitting where \dmm\ 
is explicitly considered as an additional fitting parameter.
The two component model is marginally favored by the statistical indicators and
we get 
\dmm\ = $-2.5 \pm 8.1_{\rm stat} \pm 6.2_{\rm sys}$ ppm. When we apply the correction to 
the wavelength dependent velocity drift we find \dmm\ = $-7.6 \pm 8.1_{\rm stat}\pm 6.3_{\rm sys}$ ppm. 
 It will be important to
check the extent to which the velocity drift we notice in this study
is present in UVES data used for previous \dmm\ measurements.
}
\end{abstract}

\section{Introduction}
Fundamental theories in physics rely on a set of free parameters whose 
values have to be determined experimentally and cannot be calculated 
on the basis of our present knowledge of physics. These free parameters are called 
fundamental constants as they are assumed to be time and space independent 
in the simpler   of the successful physical theories 
\citep[see][and references therein]{Uzan11a}. 
The fine structure constant, $\alpha  \equiv e^2$ / $\hbar c$, and the 
proton-to-electron mass ratio, $\mu$, are two such  dimensionless constants 
that are more straightforward to be measured experimentally. 
Current laboratory  measurements exclude any significant variation 
of these dimensionless constants over solar system scales and on geological 
time scales \citep[see][]{Olive04,Petrov06,Rosenband08,Shelkovnikov08}. 
However, it is neither observationally nor experimentally excluded that 
these fundamental constants could vary over cosmological distances and time scales. 
Therefore, constraining the temporal and spatial variations of these constants 
can have a direct impact on cosmology and fundamental physics \citep{Amendola12,Ferreira12}. 

It is known that the wavelengths of the rovibronic  
molecular transitions are sensitive to $\mu$. 
In a diatomic molecule the energy of the rotational transitions 
is proportional to the reduced mass 
of the molecule, $M$, and that of vibrational transitions is proportional  
to $\sqrt{M}$, in the first order approximation. The frequency of 
the rovibronic transitions in Born-Oppenheimer approximation 
can be written as,
\begin{equation}
\nu = c_{\rm elec} + c_{\rm vib} / \sqrt{\mu} + c_{\rm rot} / \mu
\end{equation}
where $c_{\rm elec}$, $c_{\rm vib}$, and $c_{\rm rot}$ are some numerical coefficients 
related, respectively, to electronic, vibrational and  rotational transitions. 
Therefore, by comparing the wavelength of the molecular transitions detected in 
quasar spectra with their laboratory values one can measure the 
variation in $\mu$ (i.e. \dmm\ $\equiv (\mu_z-\mu_0)/\mu_0$ where $\mu_z$ and $\mu_0$ are 
the values of proton-to-electron mass ratio at redshift $z$ and today) over cosmological time scales. 
Using  intervening molecular absorption lines seen in the high-$z$
quasar spectra for measuring  \dmm\ in the distant universe was first proposed 
by \citet{Thompson75}. As H$_2$ is the most abundant 
molecule its Lyman and Werner absorption lines seen in the quasar absorption 
spectra have been frequently used to constrain the variation of $\mu$.
However, H$_2$ molecules are 
detected  in only a few percent of
the high redshift damped Lyman-$\alpha$ (DLA) systems 
\citep{Petitjean00,ledoux03,Noterdaeme08,Srianand12,Jorgenson13}
with only a handful 
of them being suitable for probing the variation of $\mu$ \citep[see][]{Petitjean09}.

If $\mu$ varies, the observed wavelengths of different H$_2$ lines will shift
differently with respect to their expected wavelengths based on laboratory
measurements and the absorption redshift. The sensitivity of the wavelength of 
the i'th H$_2$ transition to the variation 
of $\mu$ is generally parametrised as
\begin{equation}
\lambda_i = \lambda_i^0 (1+z_{\rm abs})\big{(}1+K_i\frac{\Delta\mu}{\mu}\big{)},
\label{eq_dm}
\end{equation}
where $\lambda_i^0$ is the rest frame wavelength of the transition, $\lambda_i$  
is the observed wavelength, $K_i$ is the sensitivity coefficient 
of i'th transition, and \zabs\ is the redshift of the H$_2$ absorber. 
Alternatively  Eq. \ref{eq_dm} can be written as 
\begin{equation}
z_i = z_{\rm abs} + C K_i, ~~~~~ C = (1+z_{\rm abs})\frac{\Delta\mu}{\mu}
\label{eq_dm_a}
\end{equation}
which clearly shows that \zabs\ is only the mean redshift of transitions with $K_i$ = 0. 
Eq. \ref{eq_dm_a}  is sometimes presented as 
\begin{equation}
z_{\rm red} \equiv \frac{(z_i - z_{\rm abs})}{(1 + z_{\rm abs})} = K_i \frac{\Delta\mu}{\mu}
\label{eq_dm_b}
\end{equation}
that shows the value of \dmm\ can be determined using a linear regression analysis of reduced 
redshift ($z_{\rm red}$) vs $K_i$. 
This method has been frequently  used in the literature for constraining the variation 
of $\mu$ \citep[see][]{Varshalovich93,Cowie95,Levshakov02mnras, Ivanchik05,Reinhold06,Ubachs07,Thompson09newa,Wendt11,Wendt12}.
However, at present  measurements of \dmm\ using \h2\ is limited 
to 6 \h2-bearing DLAs at $z \ge$ 2. All of these analyses suggest that
$|$\dmm$|\le10^{-5}$ at $2\le z \le 3$. The best reported constraints
based on a single system being \dmm\ = $+(0.3\pm3.7)\times 10^{-6}$
reported by \citet{King11} towards Q~0528$-$250. 

At $z\le 1.0$ a stringent constraint on \dmm\ is obtained using inversion transitions
of NH$_3$ and rotational molecular transitions \citep{Murphy08,Henkel09,Kanekar11}. 
The best reported limit using this technique is \dmm\ = $-(3.5\pm1.2)\times 10^{-7}$ \citep{Kanekar11}. 
\citet{Bagdonaite13} obtained the strongest
constraint to date of \dmm\ = $(0.0\pm1.0)\times10^{-7}$ at $z = 0.89$
using methanol transitions. However, \dmm\ measurements using NH$_3$ and
CH$_3$OH are restricted to only two specific systems at $z\le 1$. Alternatively
one can place good constraints using 21-cm absorption in conjunction
with metal lines and assuming all other constants have not changed
\citep[see for example][]{Tzanavaris07}. \citet{Rahmani12} have
obtained \dmm = $(0.0\pm1.50)\times 10^{-6}$ using a well selected sample
of four 21-cm absorbers at \zabs $\sim$1.3. \citet{Srianand10} have 
obtained \dmm = $(-1.7\pm1.7)\times10^{-6}$ at $z \sim$3.17 using 
the 21-cm absorber towards J1337$+$3152. 
However, one of the main systematic uncertainties in this method comes from 
how one associates 21-cm and optical absorption components. 
More robust estimates can be obtained from observations
of microwave and submillimeter molecular transitions of
the same molecule which have different sensitivities to
$\mu$-variations \citep[for a review, see][]{Kozlov13}.

%
Here we report a detailed analysis of \h2 absorption in $z$=2.4018 DLA
towards \sys\ \citep{Noterdaeme07lf} using the  Ultraviolet and Visual Echelle 
Spectrograph mounted on the Very Large Telescope (VLT/UVES)  spectra taken as part of
the UVES large programme for testing the fundamental physics \citep{Molaro13}. 


\section{observation and data reduction}\label{obs_red}
\sys\ (UM 664) with a redshift $z_{\rm em} = 2.55$ and an r-band  magnitude of 18.05 was discovered by  \citet{MacAlpine82} as 
part of their search for high redshift quasars. 
The optical spectroscopic observations of \sys\ were carried out 
using  VLT/UVES   \citep[][]{Dekker00}  Unit Telescope (UT2) 8.2-m telescope at 
Paranal (Chile) [as part of ESO Large Programme 185.A-0745 
``{\it The UVES Large Program for testing Fundamental Physics}'' \citep{Molaro13}]. 
All observations were performed using the standard beam splitter with the 
dichroic $\#2$ (setting 390+580) that covers roughly from 330 nm to 450 nm 
on the BLUE CCD and from 465 nm to 578 nm and from 583 nm to 680 nm on 
the two RED CCDs.  A slit width of 0.8$^{\prime\prime}$ and CCD readout with no 
binning were used for all the observations, resulting in a pixel size of 
$\approx$ 1.3 - 1.5 \kms\ on the BLUE CCD and spectral resolution of 
$\approx$ 60,000. All the exposures were taken with the slit  aligned 
with the parallactic angle to minimize the atmospheric dispersion effects. 
The observations are comprised of 19 exposures totalling  33.3 hours of  exposure 
time in three different observing cycles started in 2010 and finished in 
2012. The amount of observing time in different cycles are, respectively,  
10.4 , 12.5 , and 10.4 hours for the first, second, and 
third cycles. Table \ref{OBS_log} summarizes the observing date and 
exposure time 
along with the seeing and airmass for all the 19 exposures divided into 
three groups based on observing cycles.   

\citet{D'Odorico00} have shown
that the resetting of the grating between an object exposure and the
ThAr calibration lamp exposure can result in an error of the order of a few hundred meters per second 
in the wavelength calibration. To minimize this effect each science exposure was
followed immediately by an attached ThAr lamp exposure. For wavelength calibration 
of each science exposure we  use the attached mode ThAr frame just taken 
after it. 
The data  were reduced using UVES Common Pipeline Library (CPL) data reduction pipeline 
release 5.3.1\footnote{http://www.eso.org/sci/facilities/paranal/instruments/uves/doc/}
using the optimal extraction method. We used $\rm 4^{th}$ order 
polynomials to find the dispersion solutions. 
The number of suitable ThAr lines used for wavelength calibration 
was usually more than 700 and the rms error was found to be in the range 40 -- 50 \ms\ 
with zero average. However this error reflects only the   calibration error at the 
observed wavelengths of the ThAr lines that are used for wavelength calibration. The 
systematic errors affecting the wavelength calibration should be measured by other 
techniques that will be discussed later in the paper.

To avoid rebinning of the pixels we use the                                   
final un-rebinned extracted spectrum of each order produced  
by CPL.  We  apply the CPL wavelength solutions to each order and merge the orders 
by implementing a weighted mean in the overlapping regions.  
All the spectra are corrected for the motion of the observatory around the 
barycenter of the Sun-Earth system. The velocity component of the 
observatory's barycentric motion towards the line of sight to the quasar was 
calculated at the exposure mid point. Conversion of air to vacuum wavelengths  
was performed using the formula given in \citet{Edlen96}.
For the co-addition of the different exposures, we interpolated the individual 
spectra and their errors to a common wavelength  
array (while conserving the pixel size) and then computed the weighted mean using 
weights estimated from the errors in each pixel.
\begin{table*}
\caption{Log of the optical spectroscopic observation of \sys\ with VLT/UVES$^{\star}$.}
\begin{center}
\begin{tabular}{ccccccccc}
\hline
\hline
 Exposure      &Observing & Starting   & Exposure    & Seeing    & Airmass& SNR & $\Delta$v \\
 identification& date    & time (UT)  &  (s)      &(arcsec)   &        &     &  (\kms) \\
  (1)          & (2) &  (3)       & (4)         &  (5)      & (6)    & (7) &  (8) \\
\hline
\\
     \multicolumn{6}{c}{cycle 1 : 2010}\\
\hline
 EXP01 &2010-07-13    &  07:59:03.56  &6250 &0.68&1.11 & 9.04   &$+$0.16 \\
 EXP02 &2010-07-15    &  07:46:52.45  &6250 &0.81&1.12 & 8.22   &$+$0.47 \\
 EXP03 &2010-08-09    &  06:33:45.04  &6250 &0.55&1.07 &11.23   &$+$0.62 \\
 EXP04 &2010-08-10    &  07:06:48.37  &6250 &0.65&1.02 & 9.13   &$+$0.36 \\
 EXP05 &2010-08-19    &  07:45:02.18  &6250 &0.82&1.01 &10.49   &$+$0.23 \\
 EXP06 &2010-10-05    &  01:22:39.57  &6250 &1.04&1.32 & 7.56   &$+$0.40 \\
     \multicolumn{6}{c}{cycle 2 : 2011}\\                        
\hline                                                           
 EXP07 &2011-10-31    &  02:30:27.36  &6400 &1.01&1.01  & 8.35  &$+$0.30 \\
 EXP08 &2011-10-31    &  04:20:34.65  &6400 &1.28&1.10  & 8.38  &$+$0.30 \\
 EXP09 &2011-11-01    &  02:10:20.43  &6400 &0.72&1.01  & 9.04  &$+$0.42 \\
 EXP10 &2011-11-02    &  01:57:51.97  &6400 &1.05&1.01  & 8.64  &$-$0.20 \\
 EXP11 &2011-11-03    &  02:03:07.21  &6400 &1.02&1.01  & 8.66  &$+$0.36 \\
 EXP12 &2011-11-04    &  00:36:35.75  &6400 &1.63&1.10  & 8.02  &$+$0.63 \\
 EXP13 &2011-11-04    &  02:33:35.34  &6700 &1.27&1.01  & 9.35  &$+$0.58\\
                                                                 
     \multicolumn{6}{c}{cycle 3 : 2012}\\                        
\hline                                                           
 EXP14 &2012-07-16    &  08:19:06.35  &6250 &0.73&1.06  & 8.21   &$+$0.56   \\
 EXP15 &2012-07-25    &  07:34:12.34  &6250 &0.66&1.07  &10.25   &$+$0.49   \\
 EXP16 &2012-08-14    &  06:22:02.87  &6250 &0.66&1.06  &10.19   &$+$0.68   \\
 EXP17 &2012-08-16    &  05:57:14.67  &6250 &0.79&1.09  & 7.73   &$+$0.80   \\
 EXP18 &2012-08-16    &  07:54:15.53  &6250 &1.00&1.01  & 6.58   &$+$0.57   \\
 EXP19 &2012-08-22    &  07:37:54.22  &6250 &0.55&1.01  & 9.93   &$+$0.35   \\
\hline
\end{tabular}
\end{center}
\begin{flushleft}
Column 5: Seeing at the beginning of the 
exposure as recorded by  Differential Image Motion Monitor (DIMM) at Paranal. Column 6: Airmass at the beginning of the 
exposure. 
Column 7: SNR calculated from a line free region in the observed wavelength range 3786-3795 \AA.
Column 8: The mean velocity difference between the clean \h2\  lines with observed 
wavelengths larger and smaller than 3650 \AA.  \\ 
$^\star$All the exposure are taken using 390+580 setting with no binning for CCD readout 
and slit aligned to the parallactic angle.\\ 
\end{flushleft}
\label{OBS_log}
\end{table*}
%
The typical SNR measured in a line free region of 3786-3795 \AA\ is 
given in the seventh column of Table \ref{OBS_log}. We notice that 
our final combined spectrum has a SNR of $\sim$ 29 at this wavelength 
interval.


\section{Systematic uncertainties in the UVES wavelength scale}
The shortcomings of the ThAr wavelength calibration of quasar spectra taken 
with VLT/UVES have already  been discussed by a number of authors 
\citep{Chand06,Levshakov06,Molaro08,Thompson09apj,Whitmore10,Agafonova11,Wendt11,
Rahmani12,Agafonova13}. 
The availability  of 19 independent spectra taken over a 3 year period allows us to 
investigate the presence of any velocity drift as a function of 
wavelength in our spectra and study its evolution with time 
before we embark on $\Delta\mu/\mu$ measurements.
 In the last column of Table  \ref{OBS_log} we give $\Delta$v, the  
velocity offset based on the mean \zabs\ of \h2\ lines detected below and 
above 3650\AA. Ideally, if \dmm = 0 we expect this to distribute randomly around zero. 
But we notice that apart from one case the values are always positive. 
Below we use cross-correlation analysis to address this in great detail.   

\subsection{Cross-correlation analysis}\label{cr_cor_ana}
%
\begin{figure} 
\centering
\includegraphics[width=0.95\hsize,bb=18 18 594 774,clip=,angle=90]{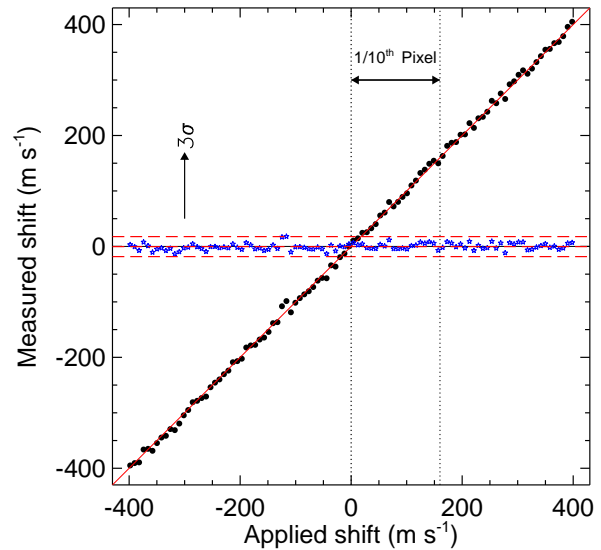}
\caption{Result of the Monte Carlo simulations to check the validity of  our cross-correlation analysis.
The abscissa is the applied shift and the ordinate is the mean of the measured 
shifts for 90 realizations. On the solid line the measured and applied shifts are identical. 
The asterisks are the residuals (measured - applied) and the long dashed lines are the 
mean and 3--$\sigma$ scatter of the residuals. The two vertical dashed lines 
indicate  1/10$^{th}$ of a pixel size ($\Delta$v $\sim$ 0.14 \kms).} 
\label{fig_sim_crcor}
\end{figure}
\begin{figure} 
\centering
\includegraphics[width=0.85\hsize,bb=18 18 594 774,clip=,angle=90]{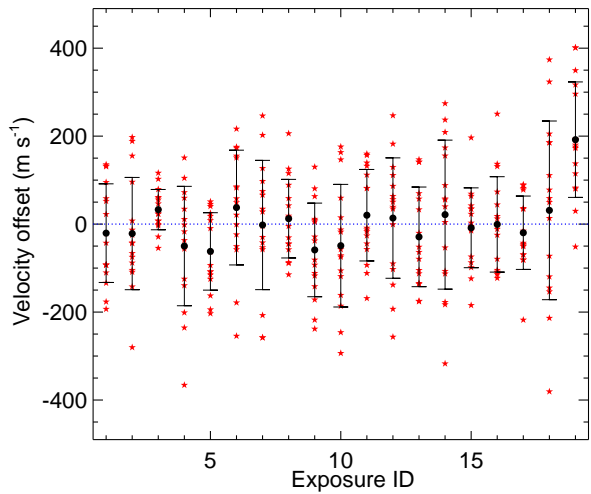}
\caption{Cross-correlation  between the spectrum of different orders in individual exposures 
 and the combined spectrum 
 (See Table \ref{tabshift_cyc1}--\ref{tabshift_cyc3} in the appendix). 
The results of cross-correlation for different orders of each 
exposure are shown as asterisks.
The mean value and 1$\sigma$ range of the shifts found for each exposure are also shown.
}
\label{fig_cross_cor0}
\end{figure}
\begin{figure} 
\centering
\includegraphics[width=0.85\hsize,bb=18 18 594 774,clip=,angle=90]{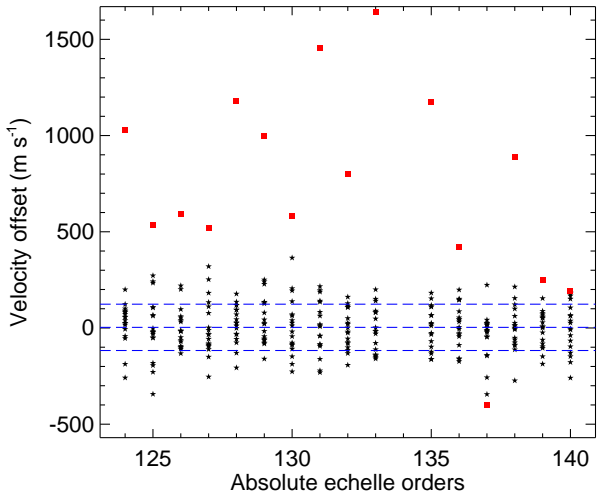}  
\caption{Cross-correlation of individual exposures with the combined spectrum after excluding EXP19. 
The results of cross-correlation for different orders of each exposure are shown as asterisks. 
Filled squares show the corresponding velocity offsets measurements for 
different regions of EXP19. The long dashed lines are marking the mean (3 \ms) and 1$\sigma$ (120 \ms) velocity scatter 
of the asterisks.} 
\label{fig_cross_cor}
\end{figure}
\begin{figure} 
\centering
\includegraphics[width=0.85\hsize,bb=18 18 594 774,clip=,angle=90]{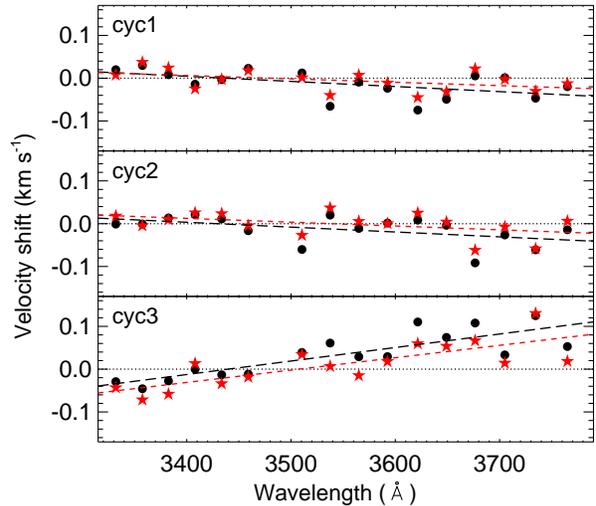}
\caption{Velocity offsets between the combined spectrum of all exposures and 
the combined spectrum for each observing cycle. The long-dashed lines 
show the line fitted to these shifts. The asterisks are the results  
after excluding EXP19 where the short-dashed lines show the line fitted to them.} 
\label{fig_cr_cor_cyc}
\end{figure}

Any systematic velocity offset that may be present between different 
spectra can be estimated using a cross-correlation technique. 
Here we cross-correlate the individual spectra as well as the combined 
spectrum of each cycle with respect to the combined spectrum of all 
19 exposures  in the wavelength windows each typically spread over  
one echelle order. As we are interested in \h2\ lines, we limit this 
analysis to 3320 $< \lambda(\AA) <$   3780 while excluding the 
wavelength range covered by the very strong \lyb\ absorption 
(i.e. echelle order number 134) of the DLA with $\log$ \nhi\  $\sim$ 21.7.  
The wavelength coverage of each window, that varies between 25--31 \AA, are 
large enough  to have a couple of saturated or 
nearly saturated absorption lines. 
This renders the cross-correlation results less sensitive to the 
 photon noise in the low SNR regions of the spectra,  thereby increasing the accuracy of such an analysis. 
We applied the cross-correlation by rebinning each pixel of size 
$\sim$ 1.4 \kms\ into 20 sub-pixels of size $\sim$ 70 \ms\ and measure 
the offset as the minimum of the $\chi^2$ curve of the flux differences 
in each window \citep[see][for more detail]{Agafonova11,Rahmani12,Wendt12,Levshakov12}. 

As our cross-correlation analysis implements a rebinning of the 
spectra on scales of 1/20 of a pixel size and involves  very fine 
interpolations it should be tested against the possible systematics 
introduced. 
To check the accuracy of our cross-correlation analysis we carried out a 
Monte Carlo simulation as follows:  (1) generate 90 realizations of the 
combined spectrum  in the range of 3750 $ < \lambda(\AA) < $ 3780, with 
a SNR roughly one fifth of the combined spectrum to mimic the individual 
exposures, (2) randomly excluding 25--35 pixels from the realized 
spectrum to  mimic the cosmic ray rejected pixels, 
(3) applying the same velocity shift to all the 90 realizations and 
(4) cross-correlating the combined spectrum with each of the 90 
spectra to measure the shifts. 
The filled circles in Fig. \ref{fig_sim_crcor} show 
the mean  measured shifts from 90 realizations vs the applied shifts 
for a sample of 100 given shifts uniformly 
chosen between -400 to 400 \ms. The residuals, shown as asterisks, 
have a standard deviation of 6 \ms\ and are  randomly distributed 
around the mean of 0 \ms. The dashed vertical lines in 
Fig. \ref{fig_sim_crcor} show the scale of one-tenth 
of our pixel size.  The exercise demonstrates that our method works 
very well in detecting the sub-pixel shifts between the combined spectra 
and the individual spectrum. 

In Fig. \ref{fig_cross_cor0} we present the measured velocity offset 
between the combined spectrum and the individual exposures in \ms\ 
over small wavelength ranges (of size typical of one echelle order).
The weighted mean and standard deviation of velocities for each 
exposure are shown as filled circle and error bar. Apart from exposure 19 (EXP19) 
all the spectra seem to have average shifts of less than 100 \ms. 
In the case of EXP19 the  average shift is 195 \ms. In addition, 
only one wavelength window has a negative shift and the rest have 
positive shifts. Therefore, this exposure seems to be severely affected 
by some systematics.  Having a reasonably high SNR, this exposure  
has already transformed this systematic error to the combined spectrum 
leading to a erroneous shift estimation. 
To have a correct estimate of the velocity offsets we make a combined 
spectrum after excluding EXP19 and repeat the cross-correlation analysis.   
Fig. \ref{fig_cross_cor} presents the amplitude of the velocity offset 
in \ms\ in different windows for all of the exposures measured with
respect to the combined spectrum excluding EXP19. The velocity offsets 
corresponding to the EXP19 are shown as filled squares. These points
have a mean value of 740 \ms. 
There also exists a mild trend for the measured  shifts of EXP19 to be 
larger in the red (smaller echelle orders) compared 
to the blue part (larger echelle orders).  We confirm the large velocity 
shifts in EXP19 using two independent  data reductions. 

The three cycles of quasar observations are separated by gaps of approximately 
one year. We next cross-correlate the combined spectrum with that of 
each cycle to check the stability of the UVES during our observations. 
The filled circles in different panels of Fig. \ref{fig_cr_cor_cyc} 
present the results of this analysis for each cycle.  
While velocity offsets of the first two cycles show a weak  
decreasing trend (top two panels of Fig.~\ref{fig_cr_cor_cyc}) with 
increasing wavelength the last cycle data shows a more pronounced trend
of velocity offset increasing with increasing wavelength.
The filled asterisks in Fig. \ref{fig_cr_cor_cyc} show 
the result of similar analysis but after excluding EXP19 both in 
the total combined spectrum and the combined spectrum of the 
third cycle.  A careful comparison of the asterisks and circles in 
the first two cycles shows that for $\lambda \ge 3500$ \AA\  all 
asterisks have larger values while for   $\lambda \le 3500$ \AA\ 
they do match. The wavelength dependent trend in the last cycle has 
been weakened a bit but still persists even after removing the 
contribution of EXP19 to the combined spectrum. 
The exercise shows that in addition to a constant shift 
there could be a wavelength dependent drift in EXP19. 
Further we also see the indication that even
other exposures taken in cycle-3 may have some systematic shift
with respect to those observed in previous 2 cycles.
Such trends if real should then be seen in the UVES spectra of the other 
objects observed in 2012. Probing this will require 
bright objects where high SNR spectrum can be obtained with short 
exposure times. 
 Asteroids have been frequently observed with UVES during 
different cycles and they provide a unique tool for this purpose. 
We test our prediction about UVES using asteroids 
in  section \ref{asteroid}. 
%

%
%
\subsection{Analysis of the asteroids spectra observed with UVES}\label{asteroid}
The cross-correlation analysis presented above allowed us to 
detect the regions and/or exposures that have large 
velocity offsets in comparison to the combined spectrum.  
However, this exercise is mainly sensitive to detect relative
shifts. 
One needs an absolute wavelength reference with very 
high accuracy  for investigating any absolute wavelength drift in 
the UVES spectrum. 
Moreover, the low SNR of the individual spectra in the wavelength range 
of $\lambda \lesssim$ 3600 \AA\ does not allow for the  direct
one-to-one comparison between different individual exposures. As a result an 
accurate understanding of the possible UVES systematics requires some other  
absolute references and/or spectra of bright objects.  
Asteroids are ideally  suited 
for this kind of analysis \citep[see][]{Molaro08} as they are 
very bright, their radial velocities are known to   an 
accuracy of 1 \ms, and their spectra are filled with the 
solar absorption features throughout any spectral range of interest.  
Several asteroids have been observed with UVES during different observing  
cycles 
for the purpose of tracking the possible wavelength calibration
issues in UVES. 
In this section we make use of the spectra of these objects observed with UVES 
to investigate the possible wavelength dependent velocity shifts during different cycles. 
To do so we select asteroids that were observed with UVES setting 
of 390 nm in the BLUE similar to our observations (see Table \ref{OBS_log}).
Table \ref{tab_ast} shows the observing log of the 
asteroids used in our analysis. We reduced these data following 
the same procedure described in section \ref{obs_red} while 
using attached mode ThAr lamp for all the exposures.  
\begin{table*}
\caption{Observing log of the UVES asteroids observations.}
\begin{center}
\begin{tabular}{lcccccc}
\hline
\hline
 Name & Observation& Exposure& Seeing  & Airmass&Spectral&Slit width\\
      &  date      &  (s)    &(arcsec) &        & resolution   & (arcsec) \\
\hline                                                                
IRIS  & 19-12-2006 & 300     & 1.21    & 1.50   & 81592    & 0.6      \\ %
      & 23-12-2006 & 300     & 1.48    & 1.47   & 82215    & 0.6      \\ %
      & 24-12-2006 & 300     & 1.25    & 1.46   & 81523    & 0.6      \\ %
      & 25-12-2006 & 300     & 1.35    & 1.44   & 81381    & 0.6      \\ %
      & 26-12-2006 & 450     & 1.79    & 1.44   & 81479    & 0.6      \\ %
      & 29-03-2012 & 600     & 1.07    & 1.14   & 59107    & 0.8      \\ %
      & 30-03-2012 & 600     & 1.20    & 1.17   & 59151    & 0.8      \\ %
      & 31-03-2012 & 600     & 1.19    & 1.18   & 58439    & 0.8      \\ %
      & 01-04-2012 & 600     & 0.99    & 1.21   & 58538    & 0.8      \\ %
CERES & 31-10-2011 & 180     & 0.99    & 1.07   & 57897    & 0.8      \\ %
      & 31-10-2011 & 180     & 0.83    & 1.09   & 57810    & 0.8      \\ %
      & 01-11-2011 & 180     & 1.09    & 1.08   & 62204    & 0.8      \\ %
      & 30-10-2010 & 180     & 1.54    & 1.37   & 60828    & 0.8      \\ %
      & 01-11-2010 & 180     & 1.20    & 1.36   & 60621    & 0.7      \\ %
      & 03-11-2010 & 180     & 0.97    & 1.41   & 60415    & 0.8      \\ %
EROS  & 27-03-2012 & 600     & 1.14    & 1.12   & 58852    & 0.8      \\ %
      & 28-03-2012 & 600     & 1.23    & 1.18   & 58535    & 0.8      \\ %
      & 29-03-2012 & 600     & 1.72    & 1.22   & 58841    & 0.8      \\ %
      & 31-03-2012 & 600     & 0.97    & 1.19   & 57681    & 0.8      \\ %
\hline
\hline
\end{tabular}
\end{center}
\label{tab_ast}
\end{table*}
Table \ref{tab_ast} shows that the time gap between different    observations can 
vary from hours to years. This allows us to probe the UVES stability in both 
short and long terms.  
We further use the solar  spectrum as an absolute reference and 
compare it with the UVES calibrated spectra of the asteroids. 
\begin{figure} 
\centering
\includegraphics[width=0.85\hsize,bb=18 18 594 774,clip=,angle=90]{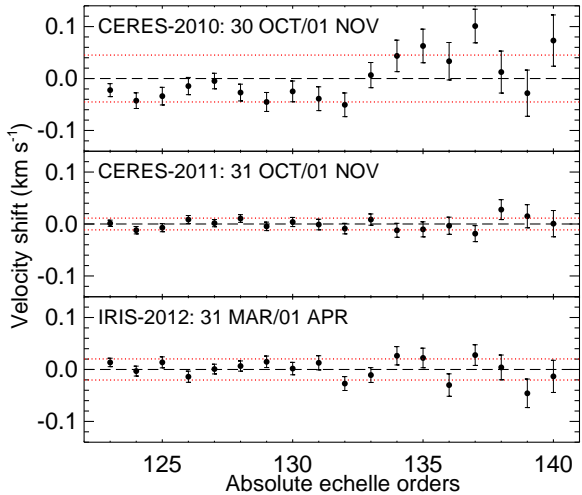}
\caption{The velocity shifts between  different asteroid exposures, after subtracting a mean velocity difference, with time gaps of one or 
two nights. The name of the 
asteroid and observing dates of the two exposures are shown in each panel. 
The standard deviation of the velocities are shown as dotted lines.} 
\label{fig_day_day}
\end{figure}
\begin{figure} 
\centering
\includegraphics[width=0.85\hsize,bb=18 18 594 774,clip=,angle=90]{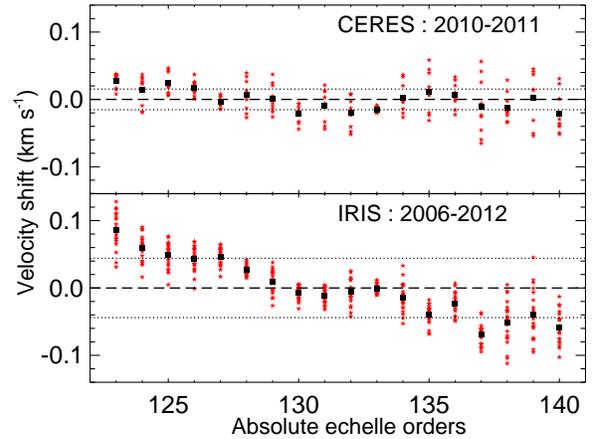}
\caption{The velocity shifts  measured in different echelle orders between asteroid exposures observed in different cycles. 
The dotted line shows the standard deviation of the velocity offsets.} 
\label{fig_cyc_cyc}
\end{figure}
%
%
\begin{figure} 
\centering
\includegraphics[width=0.85\hsize,bb=18 18 594 774,clip=,angle=90]{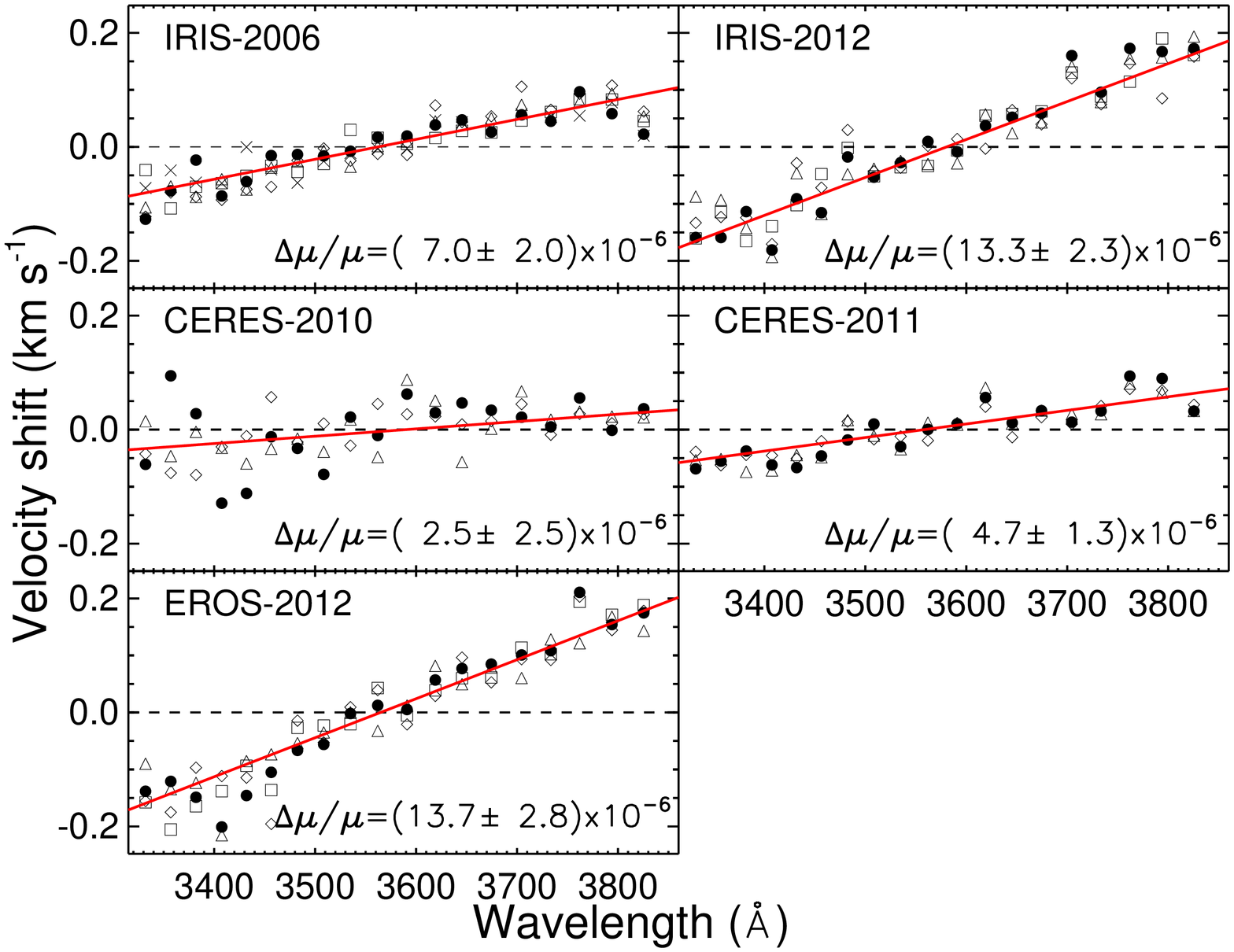}
\caption{The velocity shift measurements using cross-correlation analysis between solar and 
asteroids spectra. The solid line in each panel shows the fitted line 
to the velocities. The \dmm\ corresponding to the slope of the fitted straight line is 
also given in each panel.}
\label{fig_sol_ast}
\end{figure}
\subsubsection{Asteroid-asteroid comparison}
We apply a  cross-correlation analysis as described in Section 
\ref{cr_cor_ana} to compare asteroid spectra observed at different 
epochs. Fig. \ref{fig_day_day} shows the mean subtracted velocity 
shifts between the asteroids spectra observed with one or two nights 
of time gap during three different cycles. The abscissa is the 
absolute echelle order of the UVES and we have only shown the results 
for the orders that cover the wavelength range of 3330-3800 \AA. 
The velocity offsets hardly reach  a peak-to-peak difference of 
50 \ms\ in the case of 2011 and 2012 observation. The scatter
we notice here is very much similar to the one found by
\citet[][]{Molaro08}. The larger velocity errors and scatter seen 
in the case of 2010 observation is related to the lower SNR of the spectra of these 
asteroids as they are observed in a high airmass (see Table \ref{tab_ast}). 
However, this exercise shows that UVES is  stable over short time scales 
 (i.e. a gap of up to 2 days). Fig. \ref{fig_cyc_cyc} shows the 
velocity offsets between the spectra of IRIS observed in 2006 and 2012 (bottom panel) 
and spectra of CERES observed in  2010 and 2011 (top panel). Asterisks 
are used to show the individual velocity offsets seen in each order 
and the filled squares show the mean of them.  While in the case of 
CERES we find a (random) pattern (within a $\Delta$v = 20 ms$^{-1}$ )
 similar to what we see in Fig. \ref{fig_day_day},  in the case 
of IRIS there exists a clear steep increase of the mean velocity offsets 
as one goes towards lower echelle orders (or longer wavelengths). 
The wavelength dependent velocity shifts seen in the case of IRIS  
is a signature of a severe systematic effect affecting the UVES spectrum 
taken in the year 2012 as suggested by our cross-correlation analysis 
of the quasar spectra (see Fig. \ref{fig_cr_cor_cyc}). 
As the experiment carried out here is relative we cannot clearly 
conclude whether the problem comes from either of the cycles or both.
However, \citet{Molaro08} did not find any 
wavelength dependent systematics while comparing its absorption wavelengths 
in IRIS spectrum taken in the year 2006 
with those of solar spectra for $\lambda \ge 4000$ \AA. 
Unraveling this problem requires a very 
accurate absolute wavelength reference. We will consider the 
solar spectrum as an absolute reference for this purpose in the next 
section for further exploring this systematic.  
\subsubsection{Solar-asteroid comparison}\label{sol-ast}
\citet{Molaro08} have used the very accurate wavelengths of the solar 
absorption lines in the literature as the absolute reference and 
compared them with the measured wavelengths of the same lines in the 
asteroid spectrum observed with UVES. Unfortunately, such an
exercise is only possible for $\lambda \ge $ 4000 \AA\ 
as the solar absorption lines are 
severely blended for shorter wavelengths. 
However using an accurately calibrated solar spectrum we can 
cross-correlate it with the asteroid spectra of different years. 
%
We use the solar spectra discussed in \citet{Kurucz05,Kurucz06} 
as the solar spectrum template\footnote{The spectrum is taken from
http://kurucz.harvard.edu/sun/fluxatlas2005/}. This spectrum is corrected for telluric lines and the 
wavelength scale
of the spectrum is corrected for the gravitational redshift ($\sim$ 0.63 \kms) and 
given in air. Therefore we used UVES spectra before applying 
air-to-vacuum conversion for the correlation analysis. The
uncertainties associated with the absolute wavelength scale of
\citet{Kurucz05} is $\sim 100$ ms$^{-1}$.
We then measure the velocity offset between the solar and 
asteroid spectra in windows of the sizes of the UVES orders 
between 3330 \AA\ to 3800 \AA. 

The results of the correlation analysis are presented in 
Fig.~\ref{fig_sol_ast} as the solar-asteroid offset velocity vs 
the wavelength. Different symbols  in each panel correspond to 
different asteroid exposures obtained within a period of couple of
days.  We have subtracted  the mean velocity offset (coming from
the radial velocity differences) in each case to bring 
the mean level to zero.  
A qualitative inspection shows that the velocity offsets 
in all cases increase as wavelengths increase though with 
different slopes for different years. Obviously the two asteroids 
spectra acquired  in 2012 show the largest slopes. 
As wavelength dependent velocity shifts can mimic a non-zero \dmm, 
it is important to translate the observed trend to an apparent \dmm. 
To estimate what is the effect 
of such a wavelength dependent systematics in our \dmm\ measurements 
we carried out the following exercise: (1) First we fit a straight line 
to the velocity offset vs wavelength to get $\Delta {\rm v}(\lambda)$, 
(2) finding the offset  $\Delta {\rm v}$ and $\sigma_{\Delta {\rm v}}$ 
at the observed wavelengths of our interested \h2\ lines and assign 
a K$_i$ to each $\Delta {\rm v}$, (3) generating  2000
Gaussian realizations of each $\Delta {\rm v}$ (or reduced redshift) 
with the scatter of $\sigma_{\Delta {\rm v}}$, (4) a \dmm\ measurements from 
reduced redshift vs K analysis for 
each of the 500 realizations, and (5) finding the mean and the scatter of the 
distribution as the systematic \dmm\ and its error. 
In Fig. \ref{fig_sol_ast} we have shown the estimated \dmm\  
for each of the asteroid data. Typical error in \dmm\ is found
to be 2.5$\times10^{-6}$. The minimum \dmm\ = (2.5$\pm$2.5)$\times10^{-6}$ 
is obtained for the case of CERES-2010 observation.  
%
%
Spectra of IRIS-2012 and EROS-2012 show trends that are 
significant at more that 4.5$\sigma$ level. These trends  
translates to a \dmm\ of (13.3$\pm$2.3)$\times10^{-6}$ and
(13.7$\pm$2.8)$\times10^{-6}$ respectively for  IRIS-2012 and EROS-2012.
This clearly confirms our finding that the UVES data acquired in 2012
has large wavelength drifts. As the amplitude of \dmm\ from the wavelength
drift noted above is close to \dmm\ we wish to detect with our \h2 lines
it is important to remove these systematics from the data. Therefore,
in what follows in addition to standard \dmm\ measurements
we also present \dmm\ measurements after correcting the redshifts of
\h2 lines using the relationship found between the velocity offset
and wavelength for the asteroid spectrum obtained closest to the 
quasar observations.
\begin{figure} 
\centering
\includegraphics[width=0.85\hsize,bb=18 18 594 774,clip=,angle=90]{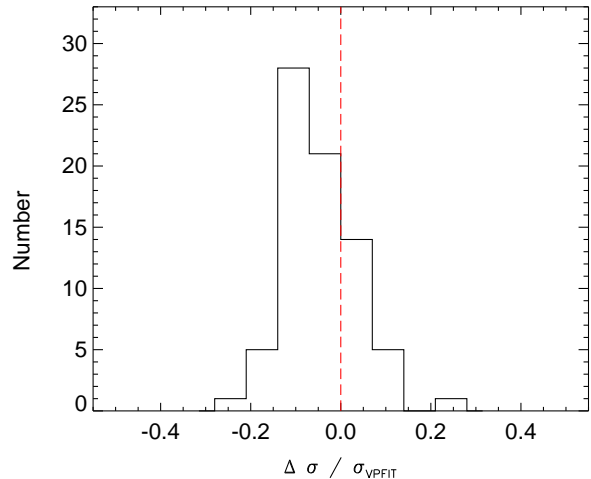}
\caption{The distribution of the relative redshift error differences between \textsc{vpfit} output and simulation. Here 
$\Delta \sigma$ = $\rm \sigma_{simulation} - \sigma_{\textsc{vpfit}}$. In 73\% of the 
simulated cases the \textsc{vpfit} 
error is larger than that of the simulation. 
This suggests that the statistical redshift  errors from \textsc{vpfit} 
are not underestimated. 
}
\label{fig_sim}
\end{figure}
\begin{figure*} 
\centering
\includegraphics[width=0.75\hsize,bb=18 18 594 774,clip=,angle=90]{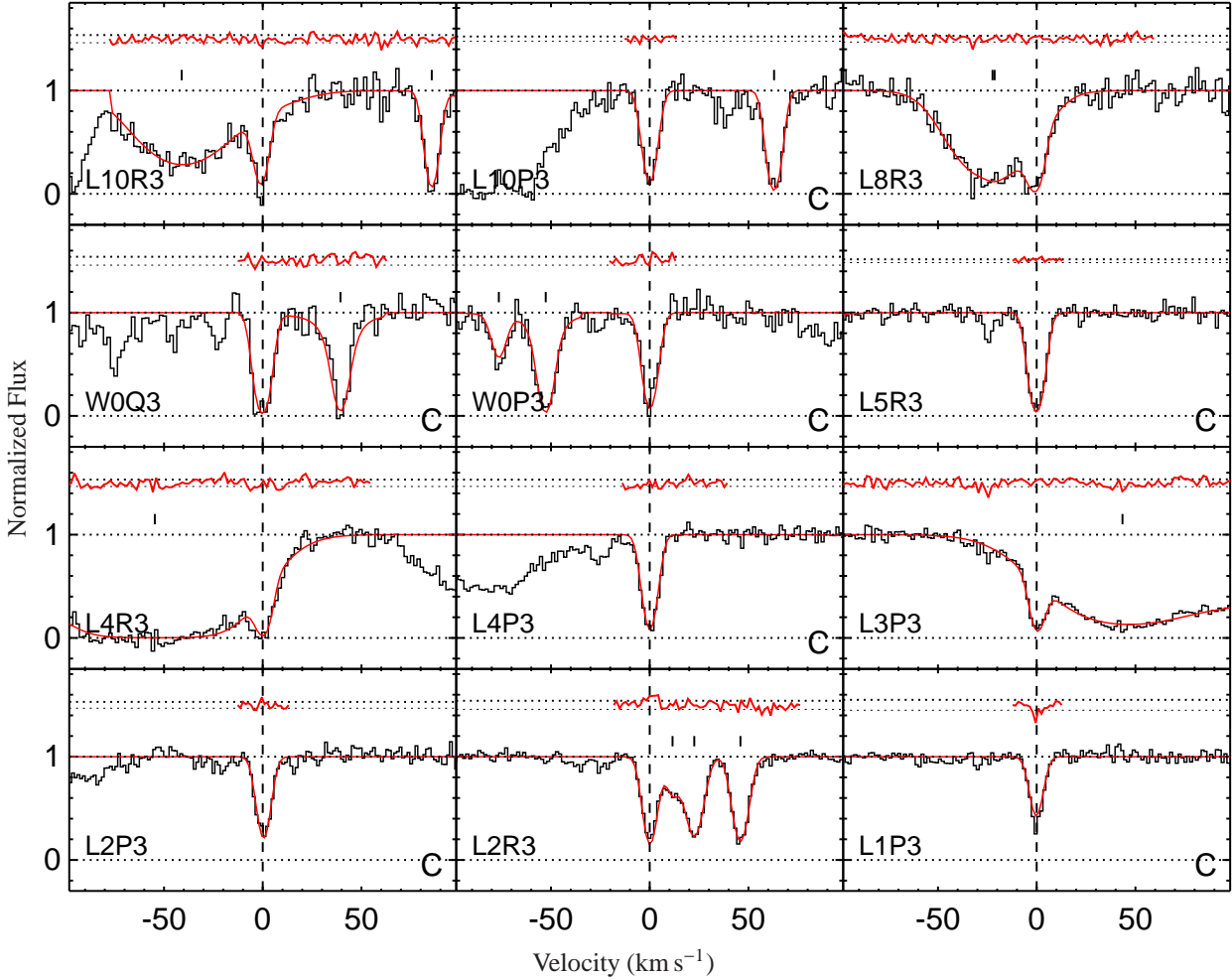}
\caption{Absorption profile of \h2\ transitions of $J$ = 3 level and the best 
fitted Voigt profile to the combined spectrum of all exposures after excluding EXP19. The normalized residual 
(i.e.([data]$-$[model])/[error]) for each 
fit is also shown in the top of each panel along with the 1$\sigma$  line. 
We  identify the clean absorption lines by  using the letter "C" in the right bottom of these transitions.  The vertical ticks mark the 
positions of fitted contamination. }
\vskip -14.0cm
\begin{picture}(400,400)(0,0)
\put( 170, 42){\large Velocity (\kms)}
\end{picture}
\vskip -14.0cm
\begin{picture}(400,400)(0,0)
\put( -29, 210){\rotatebox{90}{\large Normalized Flux}}
\end{picture}
\label{fig_J3}
\end{figure*}

In summary the analysis presented in this section suggests that one of the exposures (EXP19) shows a systematically 
large shift compared to the rest of the data. Therefore, we exclude this exposure when we discuss our final 
combined data to measure \dmm. Our analysis also suggests the existence of wavelength dependent velocity shift 
in particular in the spectra acquired in year 2012.  Therefore, we present our results for combined data of only first two 
cycles of data to gauge the influence of wavelength dependence drift in the data acquired in 2012. 
\section{Constraints on \dmm}
The $z = $ 2.4018 DLA towards \sys\ produces  more than 100 \h2\ 
absorption lines in the  observed wavelength range of 3330 \AA\ to 3800 \AA\
\citep[see Figs. 19 -- 25 in][]{Noterdaeme07lf}. These are from different 
rotational states spanning from $J =$ 0 to $J =$ 6. 
While we detect a couple of transitions of   $J =$ 6 in absorption, 
they are too weak to lead to  any reasonable estimation of the absorption 
line parameters (in particular the accurate value of $z$) of this level. Therefore,  
we do not make use of them for constraining  \dmm.  
For each rotational level we detect several absorption lines arising from 
transitions having wide range of oscillator strengths. This makes it possible 
to have very reliable estimation of fitting parameters and associated errors 
in our Voigt profile analysis of \h2\ lines.  From all the detected 
\h2\ absorption we select 71 useful lines for constraining 
\dmm, out of which 24 may have a mild contamination in the wings
from other unrelated absorption features. These mildly contaminated 
\h2\ lines are also included in the analysis as  their line centroids 
are well defined and the additional contaminations  can be modeled 
accurately through multiple-component Voigt-profile fitting.  \h2\ lines 
used in the current study and results of the single component fit are 
presented in Tables \ref{fitting_res1} and \ref{fitting_res2}. 

\citet[][]{Noterdaeme07lf}, have found that the width of  high $J$  
lines are systematically broader than that of low $J$ lines when a single 
Voigt profile component was used. As our combined spectrum
has a better SNR and pixel sampling compared to that used in 
\citet[][]{Noterdaeme07lf}, we revisit the Voigt profile fitting, using
\textsc{vpfit}\footnote{http://www.ast.cam.ac.uk/~rfc/vpfit.html},  
before measuring \dmm. First we fit all the identified
\h2\ transitions considering single $b$ and $z$ for all the levels and 
with the column density being different for different $J$-levels. Our best 
fitted model has a reduced $\chi^2$ of 1.421. The best fitted value of the 
$b$-parameter is 2.10$\pm$0.04 \kms. In addition, the derived column 
densities suggest an ortho-to-para ratio \citep[OPR; see Eq.~1 of]
[for the definition]{Srianand05} of 11.55$\pm$1.42, while $\le 3$ is expected in 
a normal local thermal equilibrium (LTE) conditions. Next we tried the fit very similar to that of
\citet[][]{Noterdaeme07lf}, where we have allowed the $b$ parameter to be 
different for different $J$-levels. In this case the best fit is obtained 
with a reduced $\chi^2$ of 1.190 and we notice that the OPR           
 is 2.26$\pm$0.15 as expected in the cold interstellar medium. 
Most of the total H$_2$ column density in this system is 
contributed by $J$ = 0 and $J$ = 1 levels. The best fitted $b$-parameters are 
0.89$\pm$0.05 \kms\ and 1.40$\pm$0.04 \kms\ respectively for $J$ = 0 and 
$J$ = 1 levels. The abnormal values of OPR obtained when we fix $b$ to be 
same for all $J$-levels can be attributed to the line saturation and 
the average $b$ being much higher than the best fitted $b$ values of $J$ = 0 and $J$ = 1.
This exercise, confirms the finding of \citet[][]{Noterdaeme07lf}
that the absorption profile of  high-$J$-levels are broader than that 
of the low-$J$ ones.
This is one of the models we use in our analysis to find the best fitted value
of \dmm. As pointed out by \citet[][]{Noterdaeme07lf}, observed differences
in the excitation temperatures and velocity width of high and low $J$-levels
may point towards multiphase nature of the absorbing gas. In order to take this
into account we allow for the mean redshift of absorption from different $J$ 
levels to be different in our analysis.  

Alternatively, one could model the two phase nature of the absorbing gas 
by using two component Voigt profile fits. In this case we constrain  $z$ and
$b$ of the two components to be the same for different $J$-levels and perform 
Voigt profile fits. Our best fit model has a reduced $\chi^2$ of 1.192 with 
two components having $b$ = 0.94$\pm$0.04 \kms\ and $b$ = 1.89$\pm$0.3 \kms and 
velocity separation of 4.4$\pm$0.2 \kms. 
This reduced $\chi^2$ is very similar to what we found for 
the single component fit with different
$b$ values for different $J$-levels discussed above.
The first component at $z = 2.401842$ contains $\sim$ 99\%
of the total \h2\ column density and has an OPR of 2.27$\pm$0.13. The second weaker
component has an OPR of 3.07$\pm$0.42. 
%
We further model the \h2\ lines with two components while allowing 
different $J$-levels to have different $b$ values. In this model $z$ is 
constrained to be same for all $J$-levels. The best fitted model in such 
a case has a reduced $\chi^2$ of 1.177  with two components having a velocity 
separation of 4.0$\pm$0.4 \kms. The reduced $\chi^2$  is slightly improved 
in comparison with the case where $b$ was tied. The first component at  
$z = 2.401844$ contains more than 99\% of the total \h2\ column density 
and has an OPR of 2.11$\pm$0.16. The second weaker component has an OPR of 2.92$\pm$0.57. 
Unlike the previous 2-component model, here both 
components have OPR consistent with what is expected under
the LTE conditions. 
We consider this two component model as the 
second case while measuring the  \dmm\ value from our data.

There are two approaches used in literature to measure \dmm\ using \h2\ absorption
lines: (i) linear regression analysis of $z_{\rm red}$ vs $K_i$ with \dmm\ as the slope 
\citep[see for example][]{Varshalovich93,Ivanchik05,Reinhold06,Wendt11}
or (ii) use \dmm\ as an additional parameter in the \textsc{vpfit} programme \citep[see for example][]{King08,Malec10,King11}. 
We employ both 
the methods to derive  \dmm\ from our data considering two cases: (i) single component
Voigt profile fit 
(called case-A)
and (ii) two component fit 
(called case-B).
 

%
\subsection{Statistical errors from \textsc{vpfit}}
%
The \textsc{vpfit} program estimates errors in each parameter
only using the diagonal terms of the covariance matrix. These are 
reliable in cases where the lines are not strongly contaminated and 
are resolved out of the instrumental resolution. 
Measured $b$ parameters of the  \h2\ lines detected towards 
\sys, especially those from $J$ = 0 and $J$ = 1 levels,  
are several times smaller than the instrumental resolution 
which is $\sim$ 5.0 \kms\ (see Table \ref{tab_nbz}). 
In such cases  the reliability 
of the \textsc{vpfit} errors should be investigated \citep[see][]{Carswell11}. 
To do so we generate 100 
simulated spectra with a same SNR as the final combined spectrum. 
For this we consider our best fitted single component Voigt profile model obtained by \textsc{vpfit} and 
add Gaussian noise  to achieve the same SNR as the original spectrum. 
We then fit the \h2\ lines 
of each mock spectrum using the same fitting regions and initial 
guess parameters as those used in case of our best fit model.   
Finally for each of the \h2\ transitions we compare the 1$\sigma$ distribution  
from the 100 mock redshifts with  
the estimated error from \textsc{vpfit}.
Fig. \ref{fig_sim} shows the distribution of the relative error differences of redshifts, 
(i.e. ($\rm \sigma_{simulation} - \sigma_{\textsc{vpfit}}) / \sigma_{\textsc{vpfit}}$) of all the \h2\ lines used 
in this analysis. Clearly when we use   majority of the transitions \textsc{vpfit} does not underestimate 
the redshift errors. As it can be seen 
from this histogram the two errors are always consistent and in $\sim$ 73\% of cases 
\textsc{vpfit} predicts a higher value for the error. We confirm that the same result
holds for  errors associated with N and $b$ parameters obtained from the
\textsc{vpfit}. We repeated the analysis for the two component fit as well and 
found that the error obtained from the \textsc{vpfit} adequately represents 
the statistical error of the parameters.
Therefore we will only use the \textsc{vpfit} errors as  statistical errors in redshifts.

\subsection{\dmm\ measurements using $z$-vs-$K$ analysis}

\begin{figure*} 
\centering
\includegraphics[width=0.7\hsize,bb=18 18 594 774,clip=,angle=90]{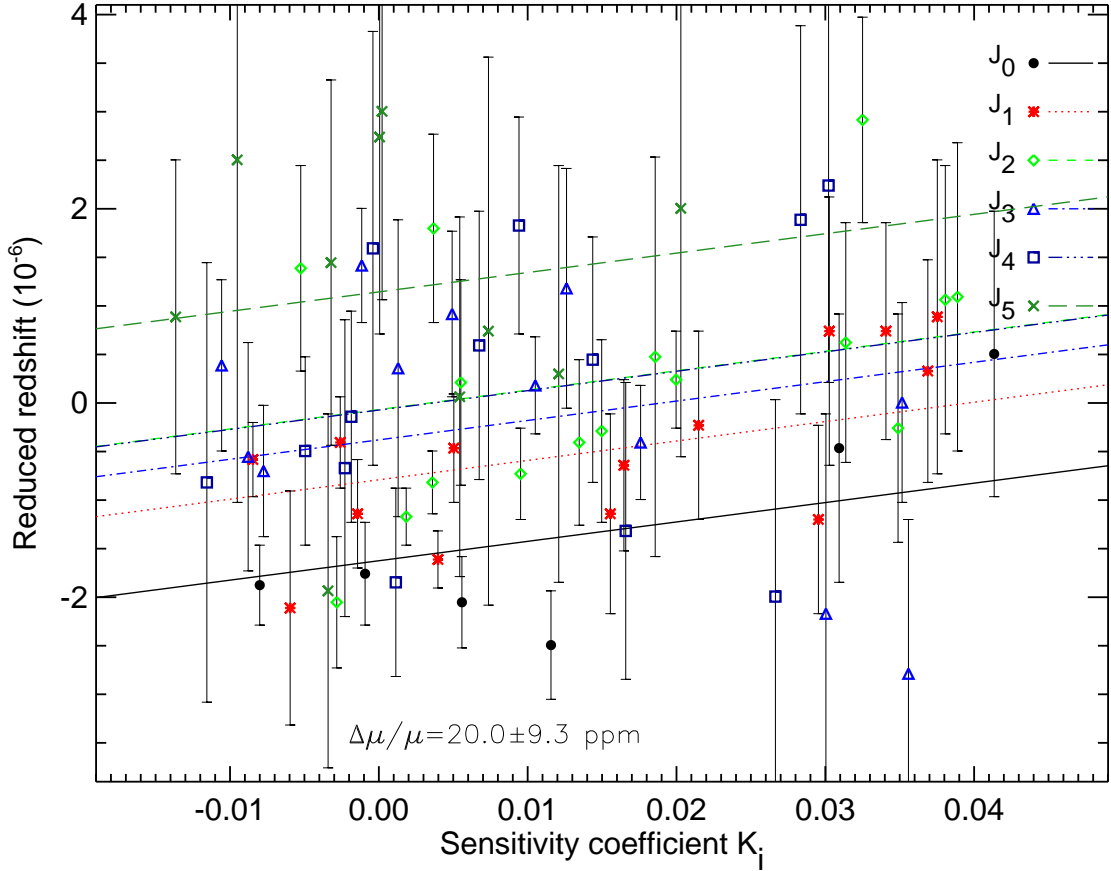}
\caption{Reduced redshift vs the  $\rm K_i$ for all the fitted \h2\ lines
in the case of combined spectrum of all exposures except EXP19. Lines
from different $J$-levels are plotted with different symbols. The best fitted
linear line for different $J$-levels with the constraint that the slope should
be same is also shown.
} 
\label{fig_zvsk_int}
\end{figure*}

Following previous studies  
\citep{Ivanchik05,Reinhold06,Ubachs07,Wendt11,Wendt12} we carry    
out \dmm\ measurements based on the  $z$-vs-$K$ linear regression
analysis in this section.
We use the redshifts of individual transitions obtained from the
\textsc{vpfit} for case-A discussed above.
Fig. \ref{fig_J3} shows our best fitted Voigt profile  for $J$ = 3 transitions 
for the combined spectrum made of all exposures excluding EXP19. 
\h2\ transitions L1P3, L2P3, L4P3, L5R3, W0Q3, and L10P3 are 
the examples of what we classify as CLEAN lines. The rest of the \h2\ 
absorption lines shown in Fig.~\ref{fig_J3} are classified as blended. 
Voigt profile fits to these lines are performed by suitably 
taking care of the blending. Fits to  \h2\ lines from other $J$-levels  
are shown in Appendix \ref{other_J}.
The fitting results for  \h2\ lines used in \dmm\ measurements are presented in  
Appendix \ref{fitting_res}. 
Table \ref{tab_nbz} summarizes the \h2\ column density, $b$ parameter,  
and the mean redshift for each $J$-level. As $J$-level increases the velocity 
offset with respect to $J$ = 0 and the $b$ parameter of the corresponding 
level also increases. The only exception is $J$ = 4 that its velocity offset and 
$b$ parameter are less than those of $J$ = 3 but still larger than $J$ = 2. 

\begin{table*}
\caption{Results of the Voigt profile analysis for different 
$J$-levels in \sys.}
\begin{center}
\begin{tabular}{ccccccc}
\hline
$J$-level&N &$z$ & $\sigma_z$&$\delta v$ &log[N(\h2$_J$)]  & $b$   \\
       & &    & \kms      & \kms  & log[cm$^{-2}$] & \kms\ \\
(1) & (2) & (3) & (4) & (5) & (6)& (7)\\
\hline
0 &6   &2.4018452&0.07&0.00 &16.91$\pm$0.02 & 0.90$\pm$0.06\\
1 &14  &2.4018486&0.05&0.30 &17.27$\pm$0.02 & 1.41$\pm$0.04\\
2 &16  &2.4018499&0.07&0.41 &14.95$\pm$0.02 & 2.68$\pm$0.07\\
3 &12  &2.4018522&0.08&0.62 &15.00$\pm$0.02 & 3.34$\pm$0.14\\
4 &13  &2.4018513&0.11&0.54 &14.19$\pm$0.02 & 2.55$\pm$0.38\\
5 &10  &2.4018550&0.15&0.87 &13.91$\pm$0.02 & 3.89$\pm$0.31\\
\hline
\end{tabular}
\end{center}
\begin{flushleft}
Column (1): indices for different rotational levels.  
Column (2): number of transitions in the given $J$-level
Column (3): mean weighted redshift of all transitions having same $J$-level.
Column (4): redshift error in \kms.
Column (5): redshift difference between the given $J$-level and $J = 0$ in \kms.
Column (6): The log of the \h2\ column density for different $J$-levels in cm$^{-2}$.
Column (7): $b$ parameter in \kms for different $J$-levels
\end{flushleft}
\label{tab_nbz}
\end{table*}

In our linear regression analysis we do not use the redshift errors
from \textsc{vpfit} as
our analysis in section \ref{asteroid} shows that the wavelengths 
of different regions may be affected by some systematic error. As 
this error is independent of the statistical redshift error  
it makes the distribution of the \h2\ redshift to have a larger scatter 
than that allowed by the statistical error we get from the Voigt profile 
fitting. Therefore, we use a bootstrap technique for estimating the 
realistic error of our regression analysis. To do so we generate 2000 
random realizations 
of the measured redshifts and estimate \dmm\ for each of these realization. We finally 
quote the 1$\sigma$ scatter of the 2000 \dmm\ as the estimated error in \dmm. 
From Table~\ref{tab_nbz} we can see that the mean redshifts of different
$J$-levels may be different in this case. Therefore in our analysis
we allow for the redshifts of different $J$ to be different by allowing
the intercept in $z$-vs-$K$ plot to be different for different $J$-levels.  
\begin{table*}
\caption{\dmm\ measurement for each cycle in $10^{-6}$ unit.}
\begin{center}
\begin{tabular}{ccc|ccccc|ccccc}
\hline
\hline
   & \multicolumn{2}{c}{z-vs-K}  & \multicolumn{10}{c}{\textsc{vpfit}}  \\ 
   & \multicolumn{2}{c}{1-component}    & \multicolumn{5}{c}{1-component}  & \multicolumn{5}{c}{2-components} \\              
(1)&(2)&(3)&(4)&(5)&(6)&(7)&(8)&(9)&(10)&(11)&(12)&(13)\\
cycle       & original & corrected$^\dagger$ & original& \chin&AICC & corrected$^\dagger$& \chin & original& \chin&AICC & corrected$^\dagger$& \chin \\
\hline
1             & $-$1.7 $\pm$16.3  & $-$4.6 $\pm$16.8 &$+$21.1$\pm$10.0&1.037&6302  &$+$19.8$\pm$ 9.9&1.029&$-$11.7$\pm$12.2 &1.032&6295  &$-$12.1$\pm$11.8&1.022\\
2             & $+$30.2$\pm$12.2  & $+$26.7$\pm$12.7 &$+$15.5$\pm$10.5&0.974&5948  &$+$10.0$\pm$10.5&0.972&$+$10.7$\pm$11.9 &0.969&5936  &$+$5.2 $\pm$11.9&0.967\\
3$^\star$     & $+$41.6$\pm$19.5  & $+$30.1$\pm$19.0 &$+$30.2$\pm$14.3&0.932&5705  &$+$14.5$\pm$12.5&0.927&$+$12.9$\pm$13.5 &0.912&5614  &$-$0.8 $\pm$13.4&0.907\\
1+2           & $+$10.7$\pm$11.4  & $+$13.8$\pm$10.2 &$+$18.8$\pm$ 7.7&1.128&6825  &$+$15.8$\pm$ 7.7&1.123&$+$0.8 $\pm$ 8.6 &1.120&6794  &$-$1.5 $\pm$ 8.7&1.115\\
1+2+3$^\star$ & $+$20.0$\pm$ 9.3  & $+$15.0$\pm$9.3  &$+$21.8$\pm$ 6.9&1.188&7167  &$+$15.6$\pm$ 6.9&1.179&$-$2.5 $\pm$ 8.1 &1.178&7115  &$-$7.6 $\pm$ 8.1&1.171\\
\hline
\hline
\end{tabular}
\end{center}
\begin{flushleft}
$^\star$ result of the cases that EXP19 is excluded.\\
$^\dagger$ results after correcting the systematics based on the solar-asteroid cross-correlation.
\end{flushleft}
\label{tab_dmu_cyc}
\end{table*}

In Fig.~\ref{fig_zvsk_int} we plot the reduced redshift vs $K$
for different transitions. The points from different $J$-levels are
marked with different symbols. The best fitted line for points
from different $J$-levels are also shown in the figure with different
line styles. As discussed before, we constrained the slope (i.e. \dmm) of
these lines to be same while allowing for the intercept (i.e. mean redshift) to
be different for different $J$-levels. The best fitted value for
\dmm\ is $20.0\pm9.3$ ppm (See column 2 of the last row in Table~\ref{tab_dmu_cyc}). 
The quoted error is obtained using bootstrapping as discussed above. 

As the wavelength dependent velocity shift is found to be minimum in the 
case of first two cycles we measured the \dmm\ using only the data
obtained in the first two cycles (i.e. 13 exposures and total integration
time of $\sim 23$ hrs). We call this sub-sample as  ``1+2''.
The results of the \dmm\ measurement for this
case is also given in Table.~\ref{tab_dmu_cyc}. We find \dmm\ = $+$10.7$\pm$11.4
ppm. As expected the mean \dmm\ obtained from this sub-sample is less
than the one obtained for the whole sample. The amount of observing time 
in different cycles are respectively  10.4 hours, 12.5 hours, and 10.4 
hours for the first, second, and third cycle. Therefore, we also measured \dmm\
using data obtained in individual cycles.
The total observing time in each cycle is sufficiently  good for 
estimating \dmm\ based on each cycle. 
We get \dmm\ = $-1.7\pm16.3$ ppm,
$+30.2\pm12.2$ ppm and $+41.6\pm19.5$ ppm respectively for the first,
second and third cycles. The progressive increase in the mean \dmm\ is 
consistent with what we notice in Fig.~\ref{fig_sol_ast} for the asteroids.

\begin{figure} 
\centering
\includegraphics[width=0.85\hsize,bb=18 18 594 774,clip=,angle=90]{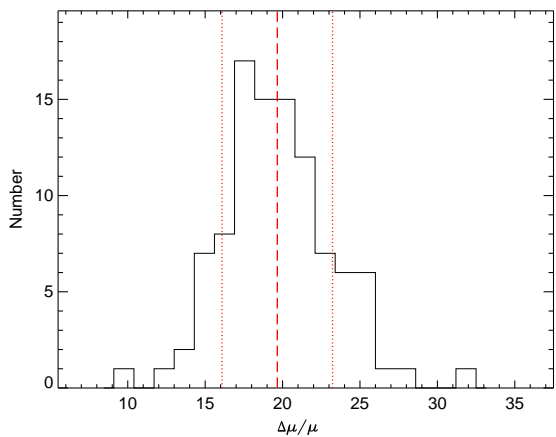}
\caption{The distribution of 100 \dmm\ measured from  
combined spectra made out of 15 randomly chosen exposures. 
The long-dashed and short-dashed lines 
show the mean and 1$\sigma$ scatter of the distribution
which respectively are 19.7$\times10^{-6}$ and 3.6$\times10^{-6}$.}
\label{fig_15_sim}
\end{figure}

The final combined spectrum used here is  based on 18 exposures.
With such a  large number of exposures, in principle the result of 
our analysis should  not be sensitive to individual exposures. 
To test this we make 100 combined spectra of 15 randomly chosen 
individual exposures and measure \dmm\ using $z$-vs-$K$ analysis of single component
fit (case-A) as discussed above for the full sample.  
Fig.  \ref{fig_15_sim} shows the distribution of the measured \dmm. 
The mean we measure (i.e. $19.7\times10^{-6}$)
is consistent with the mean we get for the full sample.
The 1$\sigma$ scatter around the mean is 3.6$\times 10^{-6}$. 
As expected this is much smaller than the statistical error 
in individual measurements. 
Thus we feel that the bootstrap method adequately quantifies 
the errors in our \dmm\ measurements.

Now we will apply corrections to the velocity drift seen in the 
case of asteroids to the individual spectrum and see its effect in
the \dmm\ measurements.
To do so we first shift the observed wavelength of each pixel 
of each individual exposure based on the modeled velocity offsets 
obtained from the solar-asteroid cross-correlation analysis in the 
corresponding observing cycle. We then combine these shifted individual
spectra to make the final combined spectrum. Only in Cycle 2 we have 
asteroid observations taken on the same nights of the quasar observation. 
In the other two cycles the nearest asteroid observations are obtained 
within 3.5 months to the quasar observations. While this is not the ideal
situation this is the best we can do. 
Results of \dmm\ measurements after applying the drift correction for different cases 
are summarized in column 3 of Table~\ref{tab_dmu_cyc}. 
The results of \dmm\ after applying drift corrections are summarized in column 4 of 
Table \ref{tab_dmu_cyc}. 
We find \dmm\ = $+15.0\pm9.3$ ppm
for the combined data after applying corrections. Clearly an offset
at the level of 5 ppm comes from this effect alone in the combined data.
We wish to note that the estimated 
\dmm\ after applying corrections should be considered as an 
indicative value as we do not have asteroid observations on
the same nights of quasar observations. In addition the quasar 
and asteroid observations are very different 
in terms of the exposure times and the source angular size. 

We notice that because of severe blending, $z$-vs-$K$ method
cannot be easily applied to the two component fit (case-B).
In the following section we obtain \dmm\ directly from
\textsc{vpfit} for both single and two component fits.

\subsection{\dmm\ measurements using \textsc{vpfit}}

We performed the Voigt profile fitting of all the chosen \h2\ lines
keeping \dmm\ as an additional fitting parameter. The results are also
summarized in columns 4 -- 13 in Table~\ref{tab_dmu_cyc}.
When we consider the single component fit (case-A) we find \dmm\ = $+21.8\pm6.9$ ppm
for the full sample with a reduced $\chi^2$ of 1.188 (see columns 4 and 5 in Table~\ref{tab_dmu_cyc}). 
This is very much consistent with what we have
found above using $z$-vs-$K$ analysis.  We find the final \dmm\ value
to be robust using different input parameter sets.  When we fit
the data obtained from first two cycles  we find \dmm\ = $+18.8\pm7.7$ ppm
with a reduced $\chi^2$ of 1.128. This also confirms our finding that 
the addition of third year data increases the measured mean of \dmm.
Table~\ref{tab_dmu_cyc} also summarizes the results of \dmm\ measurements 
for data taken on individual cycles. 
When we use the corrected spectrum for the full sample we get
\dmm\ = $+15.6\pm6.9$ ppm.  
The \dmm\ measurements for different cases after applying the correction and the 
corresponding reduced $\chi^2$ are given in columns 7 and 8 respectively. 
As noted earlier the statistical
errors from the \textsc{vpfit} are about 25 to 30\% underestimated compared to
the bootstrap errors obtained in the $z$-vs-$K$ analysis.
It is also clear from the table that correcting the velocity 
offset leads to the reduction of the \dmm\ up to 6.2 ppm for the 
combined dataset. 
Column 6 in Table~\ref{tab_dmu_cyc} gives the Akaike information criteria \citep[AIC;][]{Akaike74} 
corrected for the finite sample size \citep[AICC;][]{Sugiura78} as given in the Eq. 4 of \citet{King11}. We 
can use AICC in addition to the reduced $\chi^2$ to discriminate between the models.

Next we consider the two component Voigt profile fits (i.e. case-B) where
we keep \dmm\ as an additional fitting parameter. The results are also
summarized in columns  9 -- 13 of Table~\ref{tab_dmu_cyc}.  
The \dmm\ measurements, associated reduced $\chi^2$ and AICC parameters for uncorrected data are 
given in columns 9, 10 and 11 respectively. Results for the corrected data are provided in columns 12 and 13. 
For the whole sample
we find \dmm\ = $-2.5\pm8.1$ ppm with a reduced $\chi^2$ of 1.177. 
The reduced $\chi^2$ in this case is slightly lower than the corresponding 
single component fit. In addition we find the difference in AICC is 52 in favour of 
two component fit (i.e. case B). 
%
%
%
Table~\ref{tab_dmu_cyc} also presents results for individual cycle data 
for the two components fit. 
%

{ The comparison of AICC given in columns 6 and 11 of the 
Table~\ref{tab_dmu_cyc} also clearly favours the two component fit (i.e. case-B). 
Therefore, we will only consider  
measurements based on two component fit in the following discussions. 
%
However, bootstrapping errors in the case of  $z$-vs-$K$  linear 
regression analysis (of single component fit) are larger and robust when 
comparing with the errors from \textsc{vpfit}. In the case of combined spectrum 
of all exposures (last row of Table~\ref{tab_dmu_cyc}) we need to quadratically add 
6.2 ppm to the \textsc{vpfit} error to get the  $z$-vs-$K$ error.  
This can be considered as typical contribution of the systematic errors. 
So we consider the two component fit results with the enhanced error in 
further discussions. }
\section{Discussion}
\begin{figure} 
\centering
\includegraphics[width=0.88\hsize,bb=18 18 594 774,clip=,angle=90]{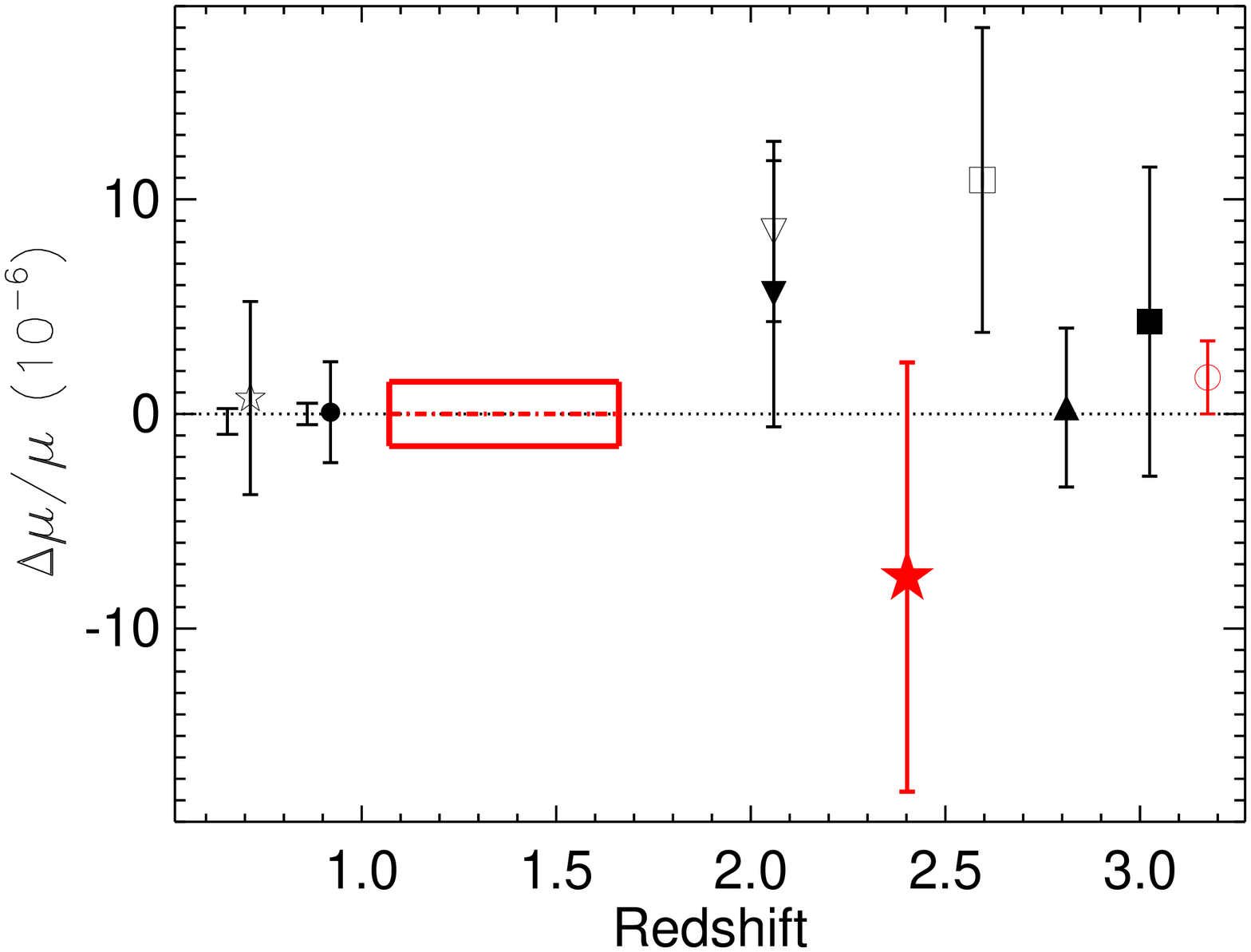}
\caption{Comparing \dmm\ measurement in this work and those in the literature. 
All measurements at 2.0 $< z < $ 3.1 are based on the analysis of \h2\ absorption. 
The filled asterisk shows our result. 
The downwards empty and filled triangles are the \dmm\ measurements 
from \citet{vanWeerdenburg11} and \citet{Malec10}. 
The  filled upward triangle and the empty and filled squares are  
respectively from \citet{King11}, \citet{King08}, 
and \citet{Wendt12}. The solid box and the open circle present the constraint obtained 
respectively by \citet{Rahmani12} and \citet{Srianand10} based on the comparison 
between 21-cm and metal lines in \MgII\ absorbers under the assumption that  $\alpha$ and $g_p$ 
have not varied. The \dmm\ at $z < 1$ are based on ammonia and methanol  inversion 
transitions that their 5$\sigma$ errors are shown. 
The two measurements at $z \sim$ 0.89 with larger and smaller errors are 
respectively from \citet[][]{Henkel09} and \citet[][]{Bagdonaite13} based on the 
same system. The two \dmm\ at $z \sim$ 0.684 with larger and smaller errors are 
respectively from \citet{Murphy08} and  \citet{Kanekar11} based on the same system. }
\label{fig_dmu_res}
\end{figure}
We have analyzed the \h2\ absorption lines from a DLA at \zabs\ = 2.4018 
towards \sys\ observed with VLT/UVES as part of the ESO Large Programme 
"The UVES large programme for testing fundamental physics". We carried out 
\dmm\  measurements based on $z$-vs-$K$ analysis. 
Our cross-correlation analysis shows that one of the exposures has  a 
large velocity shift with respect to the remaining  exposures. 
Excluding this exposure from the combined spectrum we find a 
\dmm\ = $(-2.5\pm8.1_{\rm stat}\pm6.2_{\rm sys})\times10^{-6}$. 

To understand the possible systematics affecting our observations we 
studied the asteroids observed with VLT/UVES in different cycles. 
Comparing the asteroids spectra and very accurate solar spectrum we 
show the existence of a wavelength dependent velocity shift with varying 
magnitude in different cycles. 
Correcting our observations for these systematics we measure \dmm\ = $(-7.6\pm8.1_{\rm stat}\pm6.2_{\rm sys})\times10^{-6}$.
 Our measurement 
is consistent with no variation in $\mu$ over the last 10.8 Gyr at a 
level of one part in 10$^{5}$. 
Our null result is consistent with  \dmm\ measurements in literature from 
analysis of different \h2-bearing sightlines \citep[][Table 1]{Thompson09apj}. 

Fig. \ref{fig_dmu_res} summarizes the \dmm\ measurements based on different approaches at different 
redshifts.  As can be seen our new measurement is also consistent with the 
more recent accurate measurements using \h2\ at $z\ge2$. 
\citet{Wendt12} found a \dmm\ = $(4.3\pm7.2)\times10^{-6}$ using the \h2\ absorber at 
$z$ = 3.025 towards Q0347$-$383. \citet{King11} and \citet{vanWeerdenburg11} used \h2\ and 
HD absorbers at respectively $z$ =  2.811 and 2.059 towards Q0528$-$250 and J2123$-$005 to 
find \dmm\ = $(0.3\pm3.2_{\rm stat}\pm1.9_{\rm sys})\times10^{-6}$ and \dmm\ = $(8.5\pm4.2)\times10^{-6}$. 
The measurement towards Q0528$-$250 is the most stringent \dmm\ measurements reported till date. However, 
large discrepancies (a factor of $\sim$ 50) in the reported N(\h2) values by  \citet{King11} and \citet{Noterdaeme08}
is a concern and its effect on \dmm\ needs to be investigated. 
\citet{King08} have found  \dmm\ = $(10.9\pm7.1)\times10^{-6}$  at $z$ = 2.595 towards 
Q0405$-$443. Using these measurements and ours we find 
the weighted mean of \dmm\ = $(4.1\pm3.3)\times10^{-6}$. If we use the measurement 
of \citet{Thompson09apj} of  \dmm\ = $(3.7\pm14)\times10^{-6}$ for the system towards 
Q0405$-$443 we get the mean value of  \dmm\ = $(3.2\pm2.7)\times10^{-6}$. However we 
wish to point out that three out of four UVES based measurements show positive values
of \dmm. 
As any wavelength dependent drift in these 
cases could bias these measurements towards positive \dmm\ (See Section \ref{sol-ast})   
we should exercise caution in quoting combined \dmm\ measurements. 


Best constraints on \dmm\ in quasar spectra are obtained using either 
NH$_3$ or CH$_3$OH \citep[][]{Murphy08,Henkel09,Kanekar11,Bagdonaite13}. 
These measurements reach a sensitivity of $10^{-7}$ in \dmm. However, 
only two systems at high redshift are used for these measurements and both at  
 $z < $ 1. Based on 21-cm absorption we have \dmm =  $(0.0\pm1.5)\times10^{-6}$
\citep[at $z \sim$ 1.3 by][]{Rahmani12} and \dmm =  $(-1.7\pm1.7)\times10^{-6}$ 
\citep[at $z \sim$ 3.2 by][]{Srianand10}. While these measurements are 
more stringent than \h2\ based measurements one has to assume no variation 
in $\alpha$ and $g_p$ to get a constraint on \dmm. Also care needs to be taken 
to minimize the systematics related to the line of sight to radio and optical 
emission being different. 
\section*{acknowledgement}
R. S. and P. P. J. gratefully acknowledge support 
from the Indo-French Centre for the Promotion of Advanced Research (Centre Franco-Indian pour la Promotion de la
Recherche Avanc\'ee) under contract No. 4304-2.
P.M. and C.J.M. acknowledge the financial support of grant PTDC/FIS/111725/2009 from FCT (Portugal). 
C.J.M. is also supported by an FCT Research Professorship, contract
reference IF/00064/2012.
The work of S.A.L. is supported by
DFG Sonderforschungsbereich SFB 676 Teilprojekt C4.
M.T.M.~thanks the Australian Research Council for \textsl{Discovery Project} grant DP110100866 which supported this work.
%
%
%
\def\aj{AJ}%
\def\actaa{Acta Astron.}%
\def\araa{ARA\&A}%
\def\apj{ApJ}%
\def\apjl{ApJ}%
\def\apjs{ApJS}%
\def\ao{Appl.~Opt.}%
\def\apss{Ap\&SS}%
\def\aap{A\&A}%
\def\aapr{A\&A~Rev.}%
\def\aaps{A\&AS}%
\def\azh{AZh}%
\def\baas{BAAS}%
\def\bac{Bull. astr. Inst. Czechosl.}%
\def\caa{Chinese Astron. Astrophys.}%
\def\cjaa{Chinese J. Astron. Astrophys.}%
\def\icarus{Icarus}%
\def\jcap{J. Cosmology Astropart. Phys.}%
\def\jrasc{JRASC}%
\def\mnras{MNRAS}%
\def\memras{MmRAS}%
\def\na{New A}%
\def\nar{New A Rev.}%
\def\pasa{PASA}%
\def\pra{Phys.~Rev.~A}%
\def\prb{Phys.~Rev.~B}%
\def\prc{Phys.~Rev.~C}%
\def\prd{Phys.~Rev.~D}%
\def\pre{Phys.~Rev.~E}%
\def\prl{Phys.~Rev.~Lett.}%
\def\pasp{PASP}%
\def\pasj{PASJ}%
\def\qjras{QJRAS}%
\def\rmxaa{Rev. Mexicana Astron. Astrofis.}%
\def\skytel{S\&T}%
\def\solphys{Sol.~Phys.}%
\def\sovast{Soviet~Ast.}%
\def\ssr{Space~Sci.~Rev.}%
\def\zap{ZAp}%
\def\nat{Nature}%
\def\iaucirc{IAU~Circ.}%
\def\aplett{Astrophys.~Lett.}%
\def\apspr{Astrophys.~Space~Phys.~Res.}%
\def\bain{Bull.~Astron.~Inst.~Netherlands}%
\def\fcp{Fund.~Cosmic~Phys.}%
\def\gca{Geochim.~Cosmochim.~Acta}%
\def\grl{Geophys.~Res.~Lett.}%
\def\jcp{J.~Chem.~Phys.}%
\def\jgr{J.~Geophys.~Res.}%
\def\jqsrt{J.~Quant.~Spec.~Radiat.~Transf.}%
\def\memsai{Mem.~Soc.~Astron.~Italiana}%
\def\nphysa{Nucl.~Phys.~A}%
\def\physrep{Phys.~Rep.}%
\def\physscr{Phys.~Scr}%
\def\planss{Planet.~Space~Sci.}%
\def\procspie{Proc.~SPIE}%
\let\astap=\aap
\let\apjlett=\apjl
\let\apjsupp=\apjs
\let\applopt=\ao
\bibliographystyle{mn}
\bibliography{mybib.bib}

\begin{thebibliography}{56}
\expandafter\ifx\csname natexlab\endcsname\relax\def\natexlab#1{#1}\fi

\bibitem[{{Agafonova} {et~al.}(2013){Agafonova}, {Levshakov}, {Reimers},
  {Hagen}, \& {Tytler}}]{Agafonova13}
{Agafonova}, I.~I., {Levshakov}, S.~A., {Reimers}, D., {Hagen}, H.-J., \&
  {Tytler}, D., 2013, \aap, 552, A83

\bibitem[{{Agafonova} {et~al.}(2011){Agafonova}, {Molaro}, {Levshakov}, \&
  {Hou}}]{Agafonova11}
{Agafonova}, I.~I., {Molaro}, P., {Levshakov}, S.~A., \& {Hou}, J.~L., 2011,
  \aap, 529, A28+

\bibitem[{{Akaike}(1974)}]{Akaike74}
{Akaike}, A., 1974, IEEE Trans. Autom. Control, 19, 716

\bibitem[{{Amendola} {et~al.}(2012){Amendola}, {Leite}, {Martins}, {Nunes},
  {Pedrosa}, \& {Seganti}}]{Amendola12}
{Amendola}, L., {Leite}, A.~C.~O., {Martins}, C.~J.~A.~P., {Nunes}, N.~J.,
  {Pedrosa}, P.~O.~J., \& {Seganti}, A., 2012, \prd, 86, 063515

\bibitem[{{Bagdonaite} {et~al.}(2013){Bagdonaite}, {Jansen}, {Henkel},
  {Bethlem}, {Menten}, \& {Ubachs}}]{Bagdonaite13}
{Bagdonaite}, J., {Jansen}, P., {Henkel}, C., {Bethlem}, H.~L., {Menten},
  K.~M., \& {Ubachs}, W., 2013, Science, 339, 46

\bibitem[{{Bailly} {et~al.}(2010){Bailly}, {Salumbides}, {Vervloet}, \&
  {Ubachs}}]{Ubachs10}
{Bailly}, D., {Salumbides}, E.~J., {Vervloet}, M., \& {Ubachs}, W., 2010,
  Molecular Physics, 108, 827

\bibitem[{{Carswell} {et~al.}(2011){Carswell}, {Jorgenson}, {Wolfe}, \&
  {Murphy}}]{Carswell11}
{Carswell}, R.~F., {Jorgenson}, R.~A., {Wolfe}, A.~M., \& {Murphy}, M.~T.,
  2011, \mnras, 411, 2319

\bibitem[{{Chand} {et~al.}(2006){Chand}, {Srianand}, {Petitjean}, {Aracil},
  {Quast}, \& {Reimers}}]{Chand06}
{Chand}, H., {Srianand}, R., {Petitjean}, P., {Aracil}, B., {Quast}, R., \&
  {Reimers}, D., 2006, \aap, 451, 45

\bibitem[{{Cowie} \& {Songaila}(1995)}]{Cowie95}
{Cowie}, L.~L. \& {Songaila}, A., 1995, \apj, 453, 596

\bibitem[{{Dekker} {et~al.}(2000){Dekker}, {D'Odorico}, {Kaufer}, {Delabre}, \&
  {Kotzlowski}}]{Dekker00}
{Dekker}, H., {D'Odorico}, S., {Kaufer}, A., {Delabre}, B., \& {Kotzlowski},
  H., 2000, in Proc. SPIE Vol. 4008, p. 534-545, Optical and IR Telescope
  Instrumentation and Detectors, Masanori Iye; Alan F. Moorwood; Eds., pp.
  534--545

\bibitem[{{D'Odorico} {et~al.}(2000){D'Odorico}, {Cristiani}, {Dekker}, {Hill},
  {Kaufer}, {Kim}, \& {Primas}}]{D'Odorico00}
{D'Odorico}, S., {Cristiani}, S., {Dekker}, H., {Hill}, V., {Kaufer}, A.,
  {Kim}, T., \& {Primas}, F., 2000, in Society of Photo-Optical Instrumentation
  Engineers (SPIE) Conference Series, Vol. 4005, Society of Photo-Optical
  Instrumentation Engineers (SPIE) Conference Series, {J.~Bergeron}, ed., pp.
  121--130

\bibitem[{{Edl{\'e}n}(1966)}]{Edlen96}
{Edl{\'e}n}, B., 1966, Metrologia, 2, 71

\bibitem[{{Ferreira} {et~al.}(2012){Ferreira}, {Juli{\~a}o}, {Martins}, \&
  {Monteiro}}]{Ferreira12}
{Ferreira}, M.~C., {Juli{\~a}o}, M.~D., {Martins}, C.~J.~A.~P., \& {Monteiro},
  A.~M.~R.~V.~L., 2012, \prd, 86, 125025

\bibitem[{{Henkel} {et~al.}(2009){Henkel}, {Menten}, {Murphy}, {Jethava},
  {Flambaum}, {Braatz}, {Muller}, {Ott}, \& {Mao}}]{Henkel09}
{Henkel}, C., {Menten}, K.~M., {Murphy}, M.~T., {et~al.}, 2009, \aap, 500, 725

\bibitem[{{Ivanchik} {et~al.}(2005){Ivanchik}, {Petitjean}, {Varshalovich},
  {Aracil}, {Srianand}, {Chand}, {Ledoux}, \& {Boiss{\'e}}}]{Ivanchik05}
{Ivanchik}, A., {Petitjean}, P., {Varshalovich}, D., {Aracil}, B., {Srianand},
  R., {Chand}, H., {Ledoux}, C., \& {Boiss{\'e}}, P., 2005, \aap, 440, 45

\bibitem[{{Jorgenson} {et~al.}(2013){Jorgenson}, {Murphy}, {Thompson}, \&
  {Carswell}}]{Jorgenson13}
{Jorgenson}, R.~A., {Murphy}, M.~T., {Thompson}, R., \& {Carswell}, R.~F.,
  2013, submitted to \mnras

\bibitem[{{Kanekar}(2011)}]{Kanekar11}
{Kanekar}, N., 2011, \apjl, 728, L12

\bibitem[{{King} {et~al.}(2011){King}, {Murphy}, {Ubachs}, \& {Webb}}]{King11}
{King}, J.~A., {Murphy}, M.~T., {Ubachs}, W., \& {Webb}, J.~K., 2011, \mnras,
  417, 3010

\bibitem[{{King} {et~al.}(2008){King}, {Webb}, {Murphy}, \&
  {Carswell}}]{King08}
{King}, J.~A., {Webb}, J.~K., {Murphy}, M.~T., \& {Carswell}, R.~F., 2008,
  Physical Review Letters, 101, 251304

\bibitem[{{Kozlov} \& {Levshakov}(2013)}]{Kozlov13}
{Kozlov}, M.~G. \& {Levshakov}, S.~A., 2013, Annalen der Physik, in press
  (arXiv: physics.atom-ph/1304.4510)

\bibitem[{{Kurucz}(2005)}]{Kurucz05}
{Kurucz}, R.~L., 2005, Memorie della Societa Astronomica Italiana Supplementi,
  8, 189

\bibitem[{{Kurucz}(2006)}]{Kurucz06}
---, 2006, ArXiv:astro-ph/0605029

\bibitem[{{Ledoux} {et~al.}(2003){Ledoux}, {Petitjean}, \&
  {Srianand}}]{ledoux03}
{Ledoux}, C., {Petitjean}, P., \& {Srianand}, R., 2003, \mnras, 346, 209

\bibitem[{{Levshakov} {et~al.}(2006){Levshakov}, {Centuri{\'o}n}, {Molaro},
  {D'Odorico}, {Reimers}, {Quast}, \& {Pollmann}}]{Levshakov06}
{Levshakov}, S.~A., {Centuri{\'o}n}, M., {Molaro}, P., {D'Odorico}, S.,
  {Reimers}, D., {Quast}, R., \& {Pollmann}, M., 2006, \aap, 449, 879

\bibitem[{{Levshakov} {et~al.}(2012){Levshakov}, {Combes}, {Boone},
  {Agafonova}, {Reimers}, \& {Kozlov}}]{Levshakov12}
{Levshakov}, S.~A., {Combes}, F., {Boone}, F., {Agafonova}, I.~I., {Reimers},
  D., \& {Kozlov}, M.~G., 2012, \aap, 540, L9

\bibitem[{{Levshakov} {et~al.}(2002){Levshakov}, {Dessauges-Zavadsky},
  {D'Odorico}, \& {Molaro}}]{Levshakov02mnras}
{Levshakov}, S.~A., {Dessauges-Zavadsky}, M., {D'Odorico}, S., \& {Molaro}, P.,
  2002, \mnras, 333, 373

\bibitem[{{MacAlpine} \& {Feldman}(1982)}]{MacAlpine82}
{MacAlpine}, G.~M. \& {Feldman}, F.~R., 1982, \apj, 261, 412

\bibitem[{{Malec} {et~al.}(2010){Malec}, {Buning}, {Murphy}, {Milutinovic},
  {Ellison}, {Prochaska}, {Kaper}, {Tumlinson}, {Carswell}, \&
  {Ubachs}}]{Malec10}
{Malec}, A.~L., {Buning}, R., {Murphy}, M.~T., {et~al.}, 2010, \mnras, 403,
  1541

\bibitem[{{Molaro} {et~al.}(2013){Molaro}, {Centuri{\'o}n}, {Whitmore},
  {Evans}, {Murphy}, {Agafonova}, {Bonifacio}, {D'Odorico}, {Levshakov},
  {Lopez}, {Martins}, {Petitjean}, {Rahmani}, {Reimers}, {Srianand}, {Vladilo},
  \& {Wendt}}]{Molaro13}
{Molaro}, P., {Centuri{\'o}n}, M., {Whitmore}, J.~B., {et~al.}, 2013, \aap,
  555, A68

\bibitem[{{Molaro} {et~al.}(2008){Molaro}, {Levshakov}, {Monai},
  {Centuri{\'o}n}, {Bonifacio}, {D'Odorico}, \& {Monaco}}]{Molaro08}
{Molaro}, P., {Levshakov}, S.~A., {Monai}, S., {Centuri{\'o}n}, M.,
  {Bonifacio}, P., {D'Odorico}, S., \& {Monaco}, L., 2008, \aap, 481, 559

\bibitem[{{Murphy} {et~al.}(2008){Murphy}, {Flambaum}, {Muller}, \&
  {Henkel}}]{Murphy08}
{Murphy}, M.~T., {Flambaum}, V.~V., {Muller}, S., \& {Henkel}, C., 2008,
  Science, 320, 1611

\bibitem[{{Noterdaeme} {et~al.}(2007){Noterdaeme}, {Ledoux}, {Petitjean}, {Le
  Petit}, {Srianand}, \& {Smette}}]{Noterdaeme07lf}
{Noterdaeme}, P., {Ledoux}, C., {Petitjean}, P., {Le Petit}, F., {Srianand},
  R., \& {Smette}, A., 2007, \aap, 474, 393

\bibitem[{{Noterdaeme} {et~al.}(2008){Noterdaeme}, {Ledoux}, {Petitjean}, \&
  {Srianand}}]{Noterdaeme08}
{Noterdaeme}, P., {Ledoux}, C., {Petitjean}, P., \& {Srianand}, R., 2008, \aap,
  481, 327

\bibitem[{Olive \& Skillman(2004)}]{Olive04}
Olive, K. \& Skillman, E., 2004, Astrophys. J., 617, 29

\bibitem[{{Petitjean} {et~al.}(2009){Petitjean}, {Srianand}, {Chand},
  {Ivanchik}, {Noterdaeme}, \& {Gupta}}]{Petitjean09}
{Petitjean}, P., {Srianand}, R., {Chand}, H., {Ivanchik}, A., {Noterdaeme}, P.,
  \& {Gupta}, N., 2009, Space Science Reviews, 35

\bibitem[{{Petitjean} {et~al.}(2000){Petitjean}, {Srianand}, \&
  {Ledoux}}]{Petitjean00}
{Petitjean}, P., {Srianand}, R., \& {Ledoux}, C., 2000, \aap, 364, L26

\bibitem[{Petrov {et~al.}(2006)Petrov, Nazarov, Onegin, Petrov, \&
  Sakhnovsky}]{Petrov06}
Petrov, Y., Nazarov, A., Onegin, M., Petrov, V., \& Sakhnovsky, E., 2006, Phys.
  Rev. C, 74

\bibitem[{{Rahmani} {et~al.}(2012){Rahmani}, {Srianand}, {Gupta}, {Petitjean},
  {Noterdaeme}, \& {V{\'a}squez}}]{Rahmani12}
{Rahmani}, H., {Srianand}, R., {Gupta}, N., {Petitjean}, P., {Noterdaeme}, P.,
  \& {V{\'a}squez}, D.~A., 2012, \mnras, 425, 556

\bibitem[{{Reinhold} {et~al.}(2006){Reinhold}, {Buning}, {Hollenstein},
  {Ivanchik}, {Petitjean}, \& {Ubachs}}]{Reinhold06}
{Reinhold}, E., {Buning}, R., {Hollenstein}, U., {Ivanchik}, A., {Petitjean},
  P., \& {Ubachs}, W., 2006, \prl, 96, 151101

\bibitem[{Rosenband {et~al.}(2008)Rosenband, Hume, Schmidt, Chou, Brusch,
  Lorini, Oskay, Drullinger, Fortier, Stalnaker, Diddams, Swann, Newbury,
  Itano, Wineland, \& Bergquist}]{Rosenband08}
Rosenband, T., Hume, D., Schmidt, P., {et~al.}, 2008, Science, 319, 1808

\bibitem[{{Shelkovnikov} {et~al.}(2008){Shelkovnikov}, {Butcher}, {Chardonnet},
  \& {Amy-Klein}}]{Shelkovnikov08}
{Shelkovnikov}, A., {Butcher}, R.~J., {Chardonnet}, C., \& {Amy-Klein}, A.,
  2008, \prl, 100, 150801

\bibitem[{{Srianand} {et~al.}(2010){Srianand}, {Gupta}, {Petitjean},
  {Noterdaeme}, \& {Ledoux}}]{Srianand10}
{Srianand}, R., {Gupta}, N., {Petitjean}, P., {Noterdaeme}, P., \& {Ledoux},
  C., 2010, \mnras, 405, 1888

\bibitem[{{Srianand} {et~al.}(2012){Srianand}, {Gupta}, {Petitjean},
  {Noterdaeme}, {Ledoux}, {Salter}, \& {Saikia}}]{Srianand12}
{Srianand}, R., {Gupta}, N., {Petitjean}, P., {Noterdaeme}, P., {Ledoux}, C.,
  {Salter}, C.~J., \& {Saikia}, D.~J., 2012, \mnras, 421, 651

\bibitem[{{Srianand} {et~al.}(2005){Srianand}, {Petitjean}, {Ledoux},
  {Ferland}, \& {Shaw}}]{Srianand05}
{Srianand}, R., {Petitjean}, P., {Ledoux}, C., {Ferland}, G., \& {Shaw}, G.,
  2005, \mnras, 362, 549

\bibitem[{{Sugiura}(1978)}]{Sugiura78}
{Sugiura}, N., 1978, Commun. Stat. A-Theor., 7, 13

\bibitem[{{Thompson}(1975)}]{Thompson75}
{Thompson}, R.~I., 1975, \aplett, 16, 3

\bibitem[{{Thompson} {et~al.}(2009{\natexlab{a}}){Thompson}, {Bechtold},
  {Black}, {Eisenstein}, {Fan}, {Kennicutt}, {Martins}, {Prochaska}, \&
  {Shirley}}]{Thompson09apj}
{Thompson}, R.~I., {Bechtold}, J., {Black}, J.~H., {et~al.},
  2009{\natexlab{a}}, \apj, 703, 1648

\bibitem[{{Thompson} {et~al.}(2009{\natexlab{b}}){Thompson}, {Bechtold},
  {Black}, \& {Martins}}]{Thompson09newa}
{Thompson}, R.~I., {Bechtold}, J., {Black}, J.~H., \& {Martins}, C.~J.~A.~P.,
  2009{\natexlab{b}}, \na, 14, 379

\bibitem[{{Tzanavaris} {et~al.}(2007){Tzanavaris}, {Murphy}, {Webb},
  {Flambaum}, \& {Curran}}]{Tzanavaris07}
{Tzanavaris}, P., {Murphy}, M.~T., {Webb}, J.~K., {Flambaum}, V.~V., \&
  {Curran}, S.~J., 2007, \mnras, 374, 634

\bibitem[{{Ubachs} {et~al.}(2007){Ubachs}, {Buning}, {Eikema}, \&
  {Reinhold}}]{Ubachs07}
{Ubachs}, W., {Buning}, R., {Eikema}, K.~S.~E., \& {Reinhold}, E., 2007,
  Journal of Molecular Spectroscopy, 241, 155

\bibitem[{{Uzan}(2011)}]{Uzan11a}
{Uzan}, J.-P., 2011, Living Reviews in Relativity, 14, 2

\bibitem[{{van Weerdenburg} {et~al.}(2011){van Weerdenburg}, {Murphy}, {Malec},
  {Kaper}, \& {Ubachs}}]{vanWeerdenburg11}
{van Weerdenburg}, F., {Murphy}, M.~T., {Malec}, A.~L., {Kaper}, L., \&
  {Ubachs}, W., 2011, Physical Review Letters, 106, 180802

\bibitem[{{Varshalovich} \& {Levshakov}(1993)}]{Varshalovich93}
{Varshalovich}, D.~A. \& {Levshakov}, S.~A., 1993, Soviet Journal of
  Experimental and Theoretical Physics Letters, 58, 237

\bibitem[{{Wendt} \& {Molaro}(2011)}]{Wendt11}
{Wendt}, M. \& {Molaro}, P., 2011, \aap, 526, A96+

\bibitem[{{Wendt} \& {Molaro}(2012)}]{Wendt12}
---, 2012, \aap, 541, A69

\bibitem[{{Whitmore} {et~al.}(2010){Whitmore}, {Murphy}, \&
  {Griest}}]{Whitmore10}
{Whitmore}, J.~B., {Murphy}, M.~T., \& {Griest}, K., 2010, \apj, 723, 89

\end{thebibliography}
\appendix
\section{Results of correlation analysis}
\begin{table*}
\caption{Measured shifts between the 6 individual exposures of \sys\ observed in 2010 and the 
combined exposure made out of all 19 exposures.}
\begin{center}
\begin{tabular}{lcccccc}
\hline
\hline
regions & EXP01  & EXP02  & EXP03  & EXP04  & EXP05  & EXP06     \\
        & (\kms) & (\kms) & (\kms) & (\kms) & (\kms) & (\kms)    \\
\hline
3319$-$3345&$+$0.13$\pm$ 0.20&$+$0.16$\pm$ 0.18&$+$0.00$\pm$ 0.09&$-$0.06$\pm$ 0.12&$+$0.05$\pm$ 0.12&$+$0.16$\pm$ 0.19\\
3345$-$3370&$+$0.05$\pm$ 0.15&$-$0.01$\pm$ 0.13&$+$0.06$\pm$ 0.07&$-$0.05$\pm$ 0.13&$+$0.05$\pm$ 0.12&$+$0.05$\pm$ 0.16\\
3370$-$3395&$-$0.13$\pm$ 0.17&$+$0.19$\pm$ 0.15&$+$0.00$\pm$ 0.07&$+$0.06$\pm$ 0.12&$+$0.01$\pm$ 0.11&$-$0.02$\pm$ 0.16\\
3395$-$3421&$-$0.04$\pm$ 0.15&$-$0.11$\pm$ 0.15&$+$0.02$\pm$ 0.07&$+$0.07$\pm$ 0.15&$+$0.02$\pm$ 0.11&$-$0.25$\pm$ 0.16\\
3421$-$3446&$+$0.10$\pm$ 0.15&$-$0.07$\pm$ 0.16&$-$0.03$\pm$ 0.07&$+$0.01$\pm$ 0.13&$-$0.06$\pm$ 0.10&$+$0.02$\pm$ 0.17\\
3446$-$3472&$+$0.13$\pm$ 0.14&$-$0.10$\pm$ 0.17&$+$0.02$\pm$ 0.08&$+$0.00$\pm$ 0.15&$+$0.02$\pm$ 0.10&$+$0.06$\pm$ 0.19\\
3498$-$3523&$-$0.02$\pm$ 0.16&$+$0.20$\pm$ 0.16&$+$0.06$\pm$ 0.10&$+$0.10$\pm$ 0.15&$-$0.11$\pm$ 0.11&$-$0.18$\pm$ 0.17\\
3523$-$3552&$-$0.18$\pm$ 0.12&$-$0.04$\pm$ 0.15&$+$0.08$\pm$ 0.09&$-$0.24$\pm$ 0.12&$-$0.12$\pm$ 0.10&$-$0.06$\pm$ 0.14\\
3552$-$3578&$+$0.02$\pm$ 0.14&$-$0.28$\pm$ 0.16&$+$0.01$\pm$ 0.08&$+$0.15$\pm$ 0.14&$-$0.11$\pm$ 0.12&$+$0.17$\pm$ 0.19\\
3578$-$3607&$-$0.09$\pm$ 0.14&$-$0.07$\pm$ 0.13&$+$0.01$\pm$ 0.07&$-$0.20$\pm$ 0.15&$+$0.04$\pm$ 0.10&$+$0.08$\pm$ 0.17\\
3607$-$3636&$-$0.09$\pm$ 0.12&$-$0.14$\pm$ 0.14&$+$0.05$\pm$ 0.08&$-$0.12$\pm$ 0.11&$-$0.20$\pm$ 0.09&$+$0.00$\pm$ 0.13\\
3636$-$3662&$-$0.19$\pm$ 0.15&$-$0.04$\pm$ 0.15&$-$0.05$\pm$ 0.10&$+$0.03$\pm$ 0.16&$-$0.09$\pm$ 0.13&$+$0.15$\pm$ 0.17\\
3662$-$3691&$-$0.11$\pm$ 0.10&$+$0.01$\pm$ 0.10&$+$0.12$\pm$ 0.06&$-$0.04$\pm$ 0.09&$-$0.16$\pm$ 0.08&$+$0.18$\pm$ 0.12\\
3691$-$3719&$-$0.09$\pm$ 0.13&$-$0.02$\pm$ 0.15&$+$0.10$\pm$ 0.09&$-$0.14$\pm$ 0.13&$-$0.12$\pm$ 0.10&$+$0.22$\pm$ 0.15\\
3719$-$3750&$+$0.14$\pm$ 0.13&$-$0.09$\pm$ 0.14&$+$0.06$\pm$ 0.08&$-$0.37$\pm$ 0.14&$-$0.01$\pm$ 0.10&$-$0.05$\pm$ 0.15\\
3750$-$3780&$+$0.06$\pm$ 0.14&$+$0.10$\pm$ 0.12&$+$0.03$\pm$ 0.08&$-$0.02$\pm$ 0.12&$-$0.19$\pm$ 0.09&$+$0.09$\pm$ 0.16\\
\hline
weighted mean and error&$-$0.03$\pm$ 0.03&$-$0.02$\pm$ 0.04&$+$0.03$\pm$ 0.02&$-$0.06$\pm$ 0.03&$-$0.07$\pm$ 0.03& $+$0.04$\pm$ 0.04\\
\multicolumn{1}{l}{weighted mean and error of 6 exposures}       &    \multicolumn{6}{c}{$-$0.01$\pm$0.01}  \\  
\hline
\end{tabular}
\end{center}
\begin{flushleft}
\end{flushleft}
\label{tabshift_cyc1}
\end{table*}
\begin{table*}
\caption{Measured shifts between the 7 individual exposures of \sys\ observed in 2011 and the 
combined exposure made out of all 19 exposures.}
\begin{center}
\begin{tabular}{lccccccc}
\hline
\hline
regions & EXP07  & EXP08  & EXP09  & EXP10  & EXP11  & EXP12   & EXP13    \\
        & (\kms) & (\kms) & (\kms) & (\kms) & (\kms) & (\kms)  & (\kms)   \\
\hline
3319$-$3345&$+$0.03$\pm$ 0.15&$+$0.12$\pm$ 0.20&$-$0.12$\pm$ 0.15&$-$0.12$\pm$ 0.18&$+$0.13$\pm$ 0.14&$-$0.14$\pm$ 0.13&$+$0.15$\pm$ 0.14\\
3345$-$3370&$+$0.00$\pm$ 0.12&$+$0.04$\pm$ 0.17&$-$0.11$\pm$ 0.14&$+$0.18$\pm$ 0.15&$+$0.08$\pm$ 0.13&$-$0.19$\pm$ 0.13&$+$0.07$\pm$ 0.12\\
3370$-$3395&$-$0.26$\pm$ 0.14&$-$0.06$\pm$ 0.17&$+$0.01$\pm$ 0.13&$+$0.06$\pm$ 0.14&$+$0.08$\pm$ 0.14&$+$0.18$\pm$ 0.14&$+$0.14$\pm$ 0.13\\
3395$-$3421&$+$0.03$\pm$ 0.14&$-$0.05$\pm$ 0.16&$-$0.04$\pm$ 0.16&$-$0.16$\pm$ 0.16&$-$0.04$\pm$ 0.14&$+$0.25$\pm$ 0.13&$-$0.02$\pm$ 0.12\\
3421$-$3446&$+$0.07$\pm$ 0.16&$+$0.13$\pm$ 0.16&$-$0.17$\pm$ 0.14&$-$0.19$\pm$ 0.13&$+$0.11$\pm$ 0.13&$+$0.13$\pm$ 0.14&$+$0.03$\pm$ 0.15\\
3446$-$3472&$+$0.07$\pm$ 0.14&$+$0.02$\pm$ 0.18&$-$0.14$\pm$ 0.14&$+$0.15$\pm$ 0.13&$-$0.11$\pm$ 0.13&$+$0.09$\pm$ 0.15&$-$0.18$\pm$ 0.16\\
3498$-$3523&$-$0.21$\pm$ 0.14&$+$0.00$\pm$ 0.18&$+$0.01$\pm$ 0.13&$+$0.01$\pm$ 0.16&$-$0.17$\pm$ 0.15&$-$0.09$\pm$ 0.16&$-$0.10$\pm$ 0.16\\
3523$-$3552&$+$0.13$\pm$ 0.13&$+$0.02$\pm$ 0.15&$+$0.08$\pm$ 0.11&$-$0.01$\pm$ 0.14&$-$0.06$\pm$ 0.13&$+$0.04$\pm$ 0.14&$-$0.14$\pm$ 0.12\\
3552$-$3578&$+$0.10$\pm$ 0.14&$-$0.04$\pm$ 0.18&$+$0.13$\pm$ 0.13&$-$0.07$\pm$ 0.13&$+$0.16$\pm$ 0.14&$-$0.26$\pm$ 0.17&$-$0.17$\pm$ 0.14\\
3578$-$3607&$+$0.25$\pm$ 0.15&$-$0.09$\pm$ 0.15&$-$0.24$\pm$ 0.13&$+$0.16$\pm$ 0.13&$-$0.01$\pm$ 0.13&$+$0.06$\pm$ 0.15&$-$0.08$\pm$ 0.14\\
3607$-$3636&$+$0.20$\pm$ 0.13&$+$0.09$\pm$ 0.14&$+$0.06$\pm$ 0.12&$-$0.06$\pm$ 0.13&$-$0.01$\pm$ 0.13&$-$0.10$\pm$ 0.13&$-$0.06$\pm$ 0.11\\
3636$-$3662&$-$0.06$\pm$ 0.15&$-$0.03$\pm$ 0.14&$+$0.01$\pm$ 0.13&$-$0.11$\pm$ 0.16&$+$0.14$\pm$ 0.14&$+$0.06$\pm$ 0.17&$-$0.05$\pm$ 0.15\\
3662$-$3691&$-$0.05$\pm$ 0.09&$-$0.09$\pm$ 0.09&$-$0.08$\pm$ 0.09&$-$0.29$\pm$ 0.10&$-$0.09$\pm$ 0.09&$+$0.11$\pm$ 0.10&$-$0.13$\pm$ 0.10\\
3691$-$3719&$-$0.03$\pm$ 0.14&$+$0.21$\pm$ 0.14&$-$0.09$\pm$ 0.12&$-$0.02$\pm$ 0.13&$-$0.03$\pm$ 0.12&$+$0.03$\pm$ 0.13&$-$0.11$\pm$ 0.13\\
3719$-$3750&$-$0.04$\pm$ 0.13&$-$0.11$\pm$ 0.14&$-$0.22$\pm$ 0.15&$-$0.25$\pm$ 0.14&$-$0.02$\pm$ 0.13&$+$0.00$\pm$ 0.14&$+$0.14$\pm$ 0.13\\
3750$-$3780&$-$0.26$\pm$ 0.14&$+$0.06$\pm$ 0.13&$-$0.03$\pm$ 0.13&$-$0.07$\pm$ 0.15&$+$0.16$\pm$ 0.14&$+$0.05$\pm$ 0.14&$+$0.06$\pm$ 0.13\\
\hline
weighted mean and error&$-$0.01$\pm$ 0.03&$+$0.00$\pm$ 0.04&$-$0.05$\pm$ 0.03&$-$0.06$\pm$ 0.03&$+$0.01$\pm$ 0.03&$+$ 0.02$\pm$ 0.03&$-$0.03$\pm$ 0.03\\
\multicolumn{1}{l}{weighted mean and error of 7 exposures}       &    \multicolumn{7}{c}{$-$0.02$\pm$0.01}  \\  
\hline
\end{tabular}
\end{center}
\begin{flushleft}
\end{flushleft}
\label{tabshift_cyc2}
\end{table*}
\begin{table*}
\caption{Measured shifts between the 6 individual exposures of \sys\ observed in 2012 and the 
combined exposure made out of all 19 exposures.}
\begin{center}
\begin{tabular}{lcccccc}
\hline
\hline
regions & EXP14  & EXP15  & EXP16  & EXP17  & EXP18  & EXP19    \\
        & (\kms) & (\kms) & (\kms) & (\kms) & (\kms) & (\kms)   \\
\hline
3319$-$3345&$-$0.32$\pm$ 0.19&$+$0.03$\pm$ 0.14&$+$0.02$\pm$ 0.11&$+$0.03$\pm$ 0.21&$-$0.14$\pm$ 0.21&$+$0.03$\pm$ 0.13\\
3345$-$3370&$-$0.05$\pm$ 0.15&$-$0.01$\pm$ 0.12&$-$0.11$\pm$ 0.10&$-$0.07$\pm$ 0.17&$-$0.21$\pm$ 0.25&$+$0.08$\pm$ 0.11\\
3370$-$3395&$-$0.11$\pm$ 0.16&$-$0.09$\pm$ 0.11&$-$0.12$\pm$ 0.11&$+$0.08$\pm$ 0.17&$+$0.01$\pm$ 0.22&$+$0.18$\pm$ 0.13\\
3395$-$3421&$+$0.06$\pm$ 0.16&$+$0.02$\pm$ 0.12&$+$0.04$\pm$ 0.10&$-$0.07$\pm$ 0.20&$-$0.38$\pm$ 0.20&$-$0.05$\pm$ 0.14\\
3421$-$3446&$+$0.09$\pm$ 0.15&$-$0.18$\pm$ 0.11&$-$0.01$\pm$ 0.13&$-$0.04$\pm$ 0.14&$+$0.19$\pm$ 0.20&$+$0.08$\pm$ 0.13\\
3446$-$3472&$-$0.18$\pm$ 0.16&$-$0.07$\pm$ 0.11&$+$0.13$\pm$ 0.11&$-$0.08$\pm$ 0.15&$+$0.06$\pm$ 0.19&$+$0.19$\pm$ 0.14\\
3498$-$3523&$-$0.18$\pm$ 0.19&$+$0.20$\pm$ 0.12&$-$0.10$\pm$ 0.11&$-$0.02$\pm$ 0.16&$+$0.08$\pm$ 0.23&$+$0.11$\pm$ 0.15\\
3523$-$3552&$+$0.17$\pm$ 0.15&$-$0.01$\pm$ 0.11&$-$0.11$\pm$ 0.12&$-$0.08$\pm$ 0.15&$+$0.11$\pm$ 0.20&$+$0.32$\pm$ 0.12\\
3552$-$3578&$+$0.04$\pm$ 0.17&$+$0.00$\pm$ 0.11&$-$0.11$\pm$ 0.12&$-$0.05$\pm$ 0.16&$+$0.05$\pm$ 0.20&$+$0.40$\pm$ 0.15\\
3578$-$3607&$+$0.16$\pm$ 0.16&$+$0.00$\pm$ 0.11&$-$0.02$\pm$ 0.11&$-$0.22$\pm$ 0.17&$+$0.37$\pm$ 0.23&$+$0.18$\pm$ 0.12\\
3607$-$3636&$+$0.21$\pm$ 0.15&$-$0.12$\pm$ 0.11&$+$0.00$\pm$ 0.12&$+$0.08$\pm$ 0.14&$+$0.19$\pm$ 0.23&$+$0.40$\pm$ 0.12\\
3636$-$3662&$+$0.06$\pm$ 0.16&$+$0.03$\pm$ 0.11&$+$0.14$\pm$ 0.14&$+$0.09$\pm$ 0.14&$-$0.15$\pm$ 0.26&$+$0.17$\pm$ 0.15\\
3662$-$3691&$+$0.24$\pm$ 0.11&$-$0.10$\pm$ 0.08&$+$0.07$\pm$ 0.09&$-$0.05$\pm$ 0.11&$+$0.32$\pm$ 0.17&$+$0.35$\pm$ 0.08\\
3691$-$3719&$-$0.11$\pm$ 0.17&$+$0.06$\pm$ 0.11&$-$0.08$\pm$ 0.13&$+$0.04$\pm$ 0.14&$-$0.07$\pm$ 0.23&$+$0.14$\pm$ 0.13\\
3719$-$3750&$+$0.27$\pm$ 0.15&$+$0.04$\pm$ 0.10&$+$0.25$\pm$ 0.13&$-$0.05$\pm$ 0.14&$+$0.21$\pm$ 0.21&$+$0.19$\pm$ 0.12\\
3750$-$3780&$+$0.00$\pm$ 0.16&$+$0.07$\pm$ 0.11&$+$0.01$\pm$ 0.13&$+$0.09$\pm$ 0.16&$-$0.12$\pm$ 0.22&$+$0.30$\pm$ 0.13\\
\hline
weighted mean and error& $+$0.05$\pm$ 0.04&$-$0.02$\pm$ 0.03&$+$0.00$\pm$ 0.03&$-$0.02$\pm$ 0.04& $+$0.05$\pm$ 0.05&$+$0.20$\pm$ 0.03\\
\multicolumn{1}{l}{weighted mean and error of 6 exposures}       &    \multicolumn{6}{c}{0.04$\pm$0.01}  \\  
\hline
\end{tabular}
\end{center}
\begin{flushleft}
\end{flushleft}
\label{tabshift_cyc3}
\end{table*}
\section{Laboratory wavelength of the chosen \h2\ line along with the best fitted redshifts 
from the Vogit profile analysis. } \label{fitting_res}
\begin{table*}
\caption{Laboratory wavelength of the set of \h2\ transitions that are fitted along with 
the best redshift and errors from Vogit profile analysis. The uncontaminated (CLEAN) \h2\ lines are 
highlighted in bold letters.}
\begin{center}
\begin{tabular}{lcccc}
\hline
\hline
Line ID & Lab wavelength$^a$ (\AA)  & Redshift  & Velocity (\kms)  & $K$ coefficient$^{b}$     \\
\hline
 {\bf L10R0}&{\bf  981.4387}&{\bf  2.401853(049)}&{\bf  $+$0.35$\pm$0.44}&{\bf $+$0.041 }\\
  {\bf L7R0}&{\bf 1012.8129}&{\bf  2.401850(047)}&{\bf  $+$0.07$\pm$0.42}&{\bf $+$0.031 }\\
  {\bf L3R0}&{\bf 1062.8821}&{\bf  2.401843(019)}&{\bf  $-$0.55$\pm$0.17}&{\bf $+$0.012 }\\
        L2R0& 1077.1387&  2.401845(016)&  $-$0.41$\pm$0.14& $+$0.006   \\
  {\bf L1R0}&{\bf 1092.1952}&{\bf  2.401846(017)}&{\bf  $-$0.32$\pm$0.16}&{\bf $-$0.001 }\\
  {\bf L0R0}&{\bf 1108.1273}&{\bf  2.401845(013)}&{\bf  $-$0.36$\pm$0.12}&{\bf $-$0.008 }\\
        L9R1&  992.0163&  2.401855(055)&  $+$0.47$\pm$0.49& $+$0.038   \\
  {\bf L9P1}&{\bf  992.8096}&{\bf  2.401853(038)}&{\bf  $+$0.30$\pm$0.34}&{\bf $+$0.037 }\\
  {\bf L8R1}&{\bf 1002.4520}&{\bf  2.401854(037)}&{\bf  $+$0.43$\pm$0.33}&{\bf $+$0.034 }\\
        W0Q1& 1009.7709&  2.401845(041)&  $-$0.43$\pm$0.36& $-$0.006   \\
        L7R1& 1013.4369&  2.401854(047)&  $+$0.43$\pm$0.42& $+$0.030   \\
  {\bf L7P1}&{\bf 1014.3272}&{\bf  2.401848(033)}&{\bf  $-$0.16$\pm$0.29}&{\bf $+$0.030 }\\
  {\bf L5R1}&{\bf 1037.1498}&{\bf  2.401851(033)}&{\bf  $+$0.14$\pm$0.29}&{\bf $+$0.021 }\\
  {\bf L4R1}&{\bf 1049.9597}&{\bf  2.401849(029)}&{\bf  $+$0.01$\pm$0.26}&{\bf $+$0.016 }\\
  {\bf L4P1}&{\bf 1051.0325}&{\bf  2.401848(034)}&{\bf  $-$0.14$\pm$0.30}&{\bf $+$0.016 }\\
        L2R1& 1077.6989&  2.401850(018)&  $+$0.07$\pm$0.17& $+$0.005   \\
  {\bf L2P1}&{\bf 1078.9254}&{\bf  2.401846(010)}&{\bf  $-$0.28$\pm$0.09}&{\bf $+$0.004 }\\
  {\bf L1R1}&{\bf 1092.7324}&{\bf  2.401848(018)}&{\bf  $-$0.14$\pm$0.16}&{\bf $-$0.001 }\\
  {\bf L1P1}&{\bf 1094.0519}&{\bf  2.401850(016)}&{\bf  $+$0.08$\pm$0.15}&{\bf $-$0.003 }\\
        L0R1& 1108.6332&  2.401850(013)&  $+$0.03$\pm$0.12& $-$0.008   \\
 {\bf L10R2}&{\bf  983.5911}&{\bf  2.401855(054)}&{\bf  $+$0.53$\pm$0.48}&{\bf $+$0.039 }\\
 {\bf L10P2}&{\bf  984.8640}&{\bf  2.401855(046)}&{\bf  $+$0.53$\pm$0.41}&{\bf $+$0.038 }\\
  {\bf W1R2}&{\bf  986.2440}&{\bf  2.401852(035)}&{\bf  $+$0.26$\pm$0.32}&{\bf $+$0.006 }\\
  {\bf W1Q2}&{\bf  987.9745}&{\bf  2.401858(032)}&{\bf  $+$0.75$\pm$0.29}&{\bf $+$0.004 }\\
        L9P2&  994.8740&  2.401851(039)&  $+$0.13$\pm$0.35& $+$0.035   \\
  {\bf L8R2}&{\bf 1003.9854}&{\bf  2.401862(035)}&{\bf  $+$1.08$\pm$0.32}&{\bf $+$0.033 }\\
  {\bf L8P2}&{\bf 1005.3931}&{\bf  2.401854(041)}&{\bf  $+$0.39$\pm$0.37}&{\bf $+$0.031 }\\
        W0R2& 1009.0249&  2.401856(035)&  $+$0.62$\pm$0.32& $-$0.005   \\
  {\bf L5R2}&{\bf 1038.6902}&{\bf  2.401852(016)}&{\bf  $+$0.27$\pm$0.15}&{\bf $+$0.020 }\\
        L5P2& 1040.3672&  2.401853(070)&  $+$0.34$\pm$0.62& $+$0.019   \\
        L4R2& 1051.4985&  2.401851(031)&  $+$0.12$\pm$0.28& $+$0.015   \\
  {\bf L4P2}&{\bf 1053.2842}&{\bf  2.401850(029)}&{\bf  $+$0.08$\pm$0.26}&{\bf $+$0.013 }\\
        L3R2& 1064.9948&  2.401849(016)&  $-$0.01$\pm$0.15& $+$0.010   \\
  {\bf L2R2}&{\bf 1079.2254}&{\bf  2.401849(010)}&{\bf  $-$0.04$\pm$0.09}&{\bf $+$0.004 }\\
  {\bf L2P2}&{\bf 1081.2660}&{\bf  2.401848(010)}&{\bf  $-$0.15$\pm$0.09}&{\bf $+$0.002 }\\
  {\bf L1R2}&{\bf 1094.2446}&{\bf  2.401845(022)}&{\bf  $-$0.41$\pm$0.20}&{\bf $-$0.003 }\\
       L10R3&  985.9628&  2.401842(053)&  $-$0.63$\pm$0.47& $+$0.036   \\
 {\bf L10P3}&{\bf  987.7688}&{\bf  2.401852(035)}&{\bf  $+$0.21$\pm$0.31}&{\bf $+$0.035 }\\
        L8R3& 1006.4141&  2.401844(070)&  $-$0.44$\pm$0.62& $+$0.030   \\
  {\bf W0Q3}&{\bf 1012.6796}&{\bf  2.401850(040)}&{\bf  $+$0.04$\pm$0.35}&{\bf $-$0.009 }\\
  {\bf W0P3}&{\bf 1014.5042}&{\bf  2.401853(029)}&{\bf  $+$0.32$\pm$0.26}&{\bf $-$0.011 }\\
  {\bf L5R3}&{\bf 1041.1588}&{\bf  2.401850(020)}&{\bf  $+$0.08$\pm$0.18}&{\bf $+$0.018 }\\
        L4R3& 1053.9761&  2.401856(042)&  $+$0.56$\pm$0.37& $+$0.013   \\
  {\bf L4P3}&{\bf 1056.4714}&{\bf  2.401852(016)}&{\bf  $+$0.26$\pm$0.15}&{\bf $+$0.011 }\\
        L3P3& 1070.1408&  2.401855(028)&  $+$0.48$\pm$0.25& $+$0.005   \\
        L2R3& 1081.7112&  2.401853(052)&  $+$0.31$\pm$0.46& $+$0.001   \\
  {\bf L2P3}&{\bf 1084.5603}&{\bf  2.401857(019)}&{\bf  $+$0.63$\pm$0.17}&{\bf $-$0.001 }\\
  {\bf L1P3}&{\bf 1099.7872}&{\bf  2.401849(023)}&{\bf  $-$0.01$\pm$0.20}&{\bf $-$0.008 }\\
        W1Q4&  992.0508&  2.401857(076)&  $+$0.68$\pm$0.67& $-$0.000   \\
        W1P4&  994.2299&  2.401849(052)&  $+$0.00$\pm$0.46& $-$0.002   \\
        L9R4&  999.2715&  2.401859(068)&  $+$0.88$\pm$0.61& $+$0.030   \\
  {\bf L9P4}&{\bf 1001.6557}&{\bf  2.401858(067)}&{\bf  $+$0.77$\pm$0.60}&{\bf $+$0.028 }\\
        L8R4& 1009.7196&  2.401845(068)&  $-$0.39$\pm$0.60& $+$0.027   \\
  {\bf L6P4}&{\bf 1035.1825}&{\bf  2.401847(052)}&{\bf  $-$0.19$\pm$0.46}&{\bf $+$0.017 }\\
  {\bf L5R4}&{\bf 1044.5433}&{\bf  2.401853(042)}&{\bf  $+$0.33$\pm$0.38}&{\bf $+$0.014 }\\
        L4R4& 1057.3807&  2.401858(038)&  $+$0.75$\pm$0.34& $+$0.009   \\
  {\bf L4P4}&{\bf 1060.5810}&{\bf  2.401854(046)}&{\bf  $+$0.38$\pm$0.41}&{\bf $+$0.007 }\\
  {\bf L3P4}&{\bf 1074.3129}&{\bf  2.401845(033)}&{\bf  $-$0.35$\pm$0.29}&{\bf $+$0.001 }\\
  {\bf L2R4}&{\bf 1085.1455}&{\bf  2.401851(037)}&{\bf  $+$0.16$\pm$0.33}&{\bf $-$0.002 }\\
  {\bf L2P4}&{\bf 1088.7954}&{\bf  2.401850(033)}&{\bf  $+$0.06$\pm$0.29}&{\bf $-$0.005 }\\
  {\bf L1P4}&{\bf 1104.0839}&{\bf  2.401849(077)}&{\bf  $-$0.04$\pm$0.68}&{\bf $-$0.012 }\\
\hline
\end{tabular}
\end{center}
\begin{flushleft}
Column (1): Name of the \h2\ fitted transitions. Column (2): The laboratory wavelengths. 
Columns (2) and (3): The best fitted redshifts for \h2\ lines and their errors. 
Column (5): Velocity offset  between the redshift of a given  \h2\ transition and the weighted mean redshift 
of the all the \h2\ lines. 
Column (6)  Sensitivity coefficient of \h2\ lines.\\
{$^a$} Wavelengths are from \citet{Ubachs10}.\\
{$^b$} $K$ coefficient are from \citet{Ubachs07}.\\
\end{flushleft}
\label{fitting_res1}
\end{table*}
\begin{table*}
\captionsetup[table]{labelformat=empty,labelsep= none,justification= justified,width=.75\textwidth,aboveskip=5pt}
\caption{Table \ref{fitting_res1} continued.}
\begin{center}
\begin{tabular}{lcccc}
\hline
\hline
Line ID & Lab wavelength (\AA)  & Redshift  & Velocity (\kms)  & K factor     \\
\hline
        W1Q5&  994.9244&  2.401857(063)&  $+$0.64$\pm$0.56& $-$0.003   \\
  {\bf W0R5}&{\bf 1014.2425}&{\bf  2.401862(065)}&{\bf  $+$1.11$\pm$0.58}&{\bf $+$0.000 }\\
  {\bf L8P5}&{\bf 1017.0043}&{\bf  2.401858(086)}&{\bf  $+$0.80$\pm$0.77}&{\bf $+$0.020 }\\
  {\bf W0Q5}&{\bf 1017.8315}&{\bf  2.401855(054)}&{\bf  $+$0.47$\pm$0.48}&{\bf $-$0.014 }\\
        L6P5& 1040.0587&  2.401853(073)&  $+$0.29$\pm$0.64& $+$0.012   \\
  {\bf L5P5}&{\bf 1052.4970}&{\bf  2.401854(095)}&{\bf  $+$0.42$\pm$0.84}&{\bf $+$0.007 }\\
  {\bf L4R5}&{\bf 1061.6972}&{\bf  2.401852(062)}&{\bf  $+$0.22$\pm$0.55}&{\bf $+$0.005 }\\
  {\bf L3R5}&{\bf 1075.2441}&{\bf  2.401861(069)}&{\bf  $+$1.02$\pm$0.61}&{\bf $+$0.000 }\\
        L3P5& 1079.4004&  2.401845(062)&  $-$0.38$\pm$0.55& $-$0.003   \\
  {\bf L2P5}&{\bf 1093.9550}&{\bf  2.401860(119)}&{\bf  $+$0.96$\pm$1.06}&{\bf $-$0.010 }\\
\hline
\end{tabular}
\end{center}
\begin{flushleft}
\end{flushleft}
\label{fitting_res2}
\end{table*}

\section{Results of Voigt profile fitting analysis for different \h2\ $J$-levels}\label{other_J}
%
%

\pagestyle{empty}
\clearpage
\begin{figure*} 
\centering
\includegraphics[width=0.80\hsize,bb=18 18 594 774,clip=,angle=90]{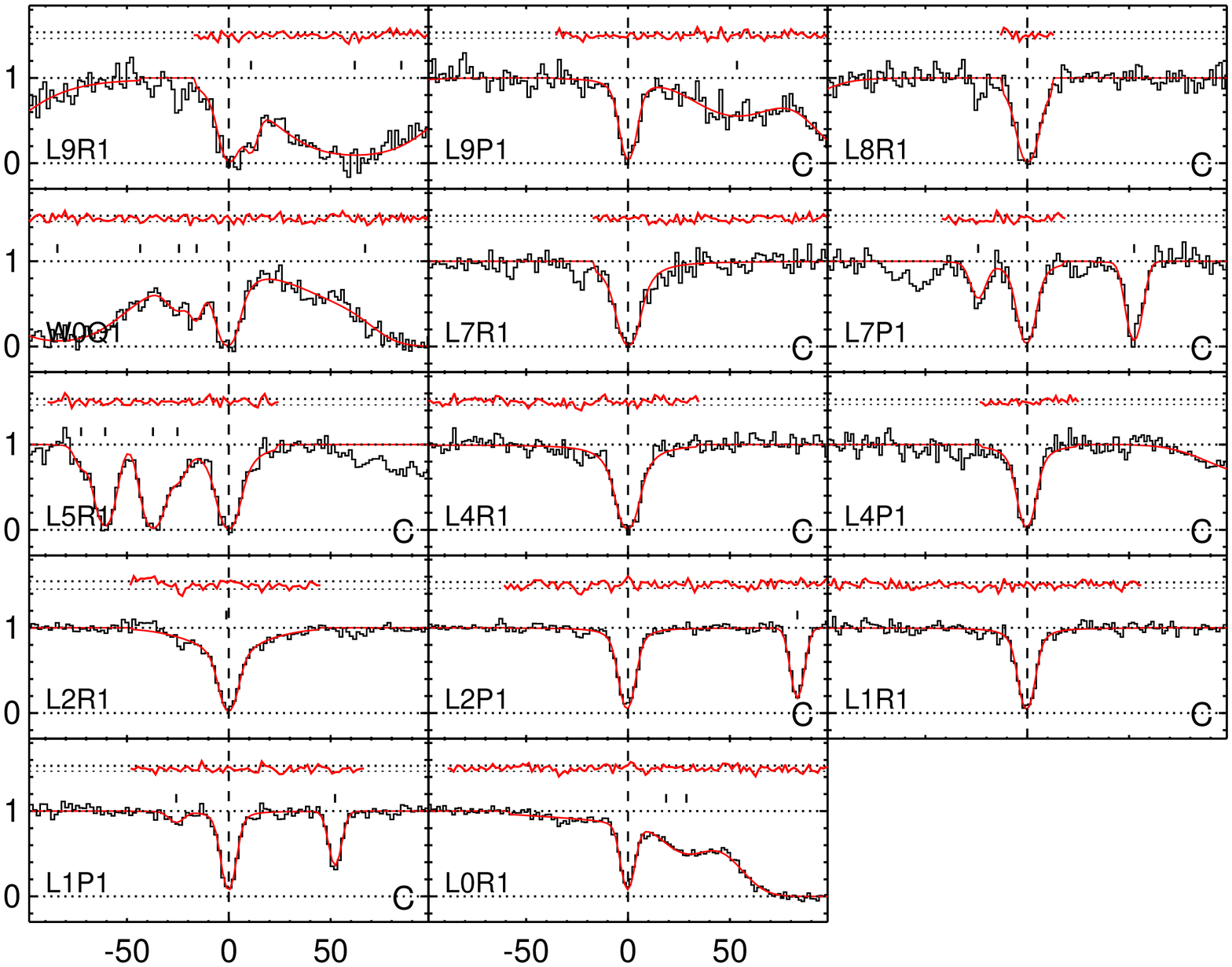}
\caption{The absorption profiles of \h2\ transitions with $J$ = 1 in \sys\ and the best fitted Voigt profile 
to the combined spectrum of all exposures after excluding EXP19. The normalized residual 
(i.e. ([data]-[model])/[error]) for each fit is shown in top of each panel along with the 
$1\sigma$  horizontal line.  The vertical ticks mark the 
positions of fitted contamination.}
\label{fig_J1}
\vskip -14.0cm
\begin{picture}(400,400)(0,0)
\put( 170, 44){\large Velocity (\kms)}
\end{picture}
\vskip -14.0cm
\begin{picture}(400,400)(0,0)
\put( -29, 225){\rotatebox{90}{\large Normalized Flux}}
\end{picture}
\label{fig_J3}
\end{figure*}%

\clearpage
\begin{figure*} 
\centering
\includegraphics[width=0.80\hsize,bb=18 18 594 774,clip=,angle=90]{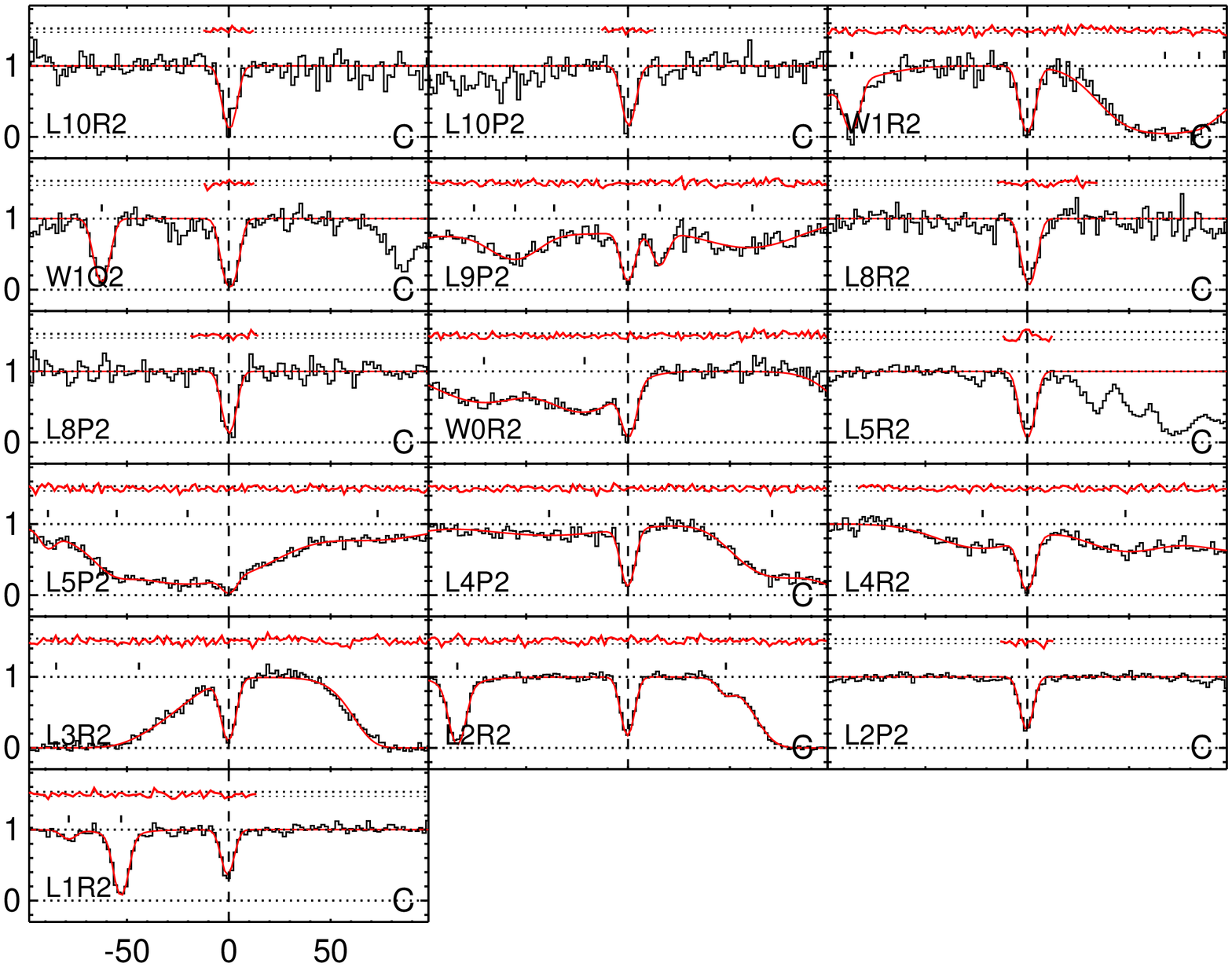}
\caption{The absorption profiles of \h2\ transitions with $J$ = 2 in \sys\ and the best fitted Voigt profile 
to the combined spectrum of all exposures after excluding EXP19. The normalized residual 
(i.e. ([data]-[model])/[error]) for each fit is shown in top of each panel along with the 
$1\sigma$  horizontal line.  The vertical ticks mark the 
positions of fitted contamination.}
\vskip -14.0cm
\begin{picture}(400,400)(0,0)
\put( 170, 44){\large Velocity (\kms)}
\end{picture}
\vskip -14.0cm
\begin{picture}(400,400)(0,0)
\put( -29, 225){\rotatebox{90}{\large Normalized Flux}}
\end{picture}

\label{fig_J2}
\end{figure*}%

\begin{figure*} 
\centering
\includegraphics[width=0.80\hsize,bb=18 18 594 774,clip=,angle=90]{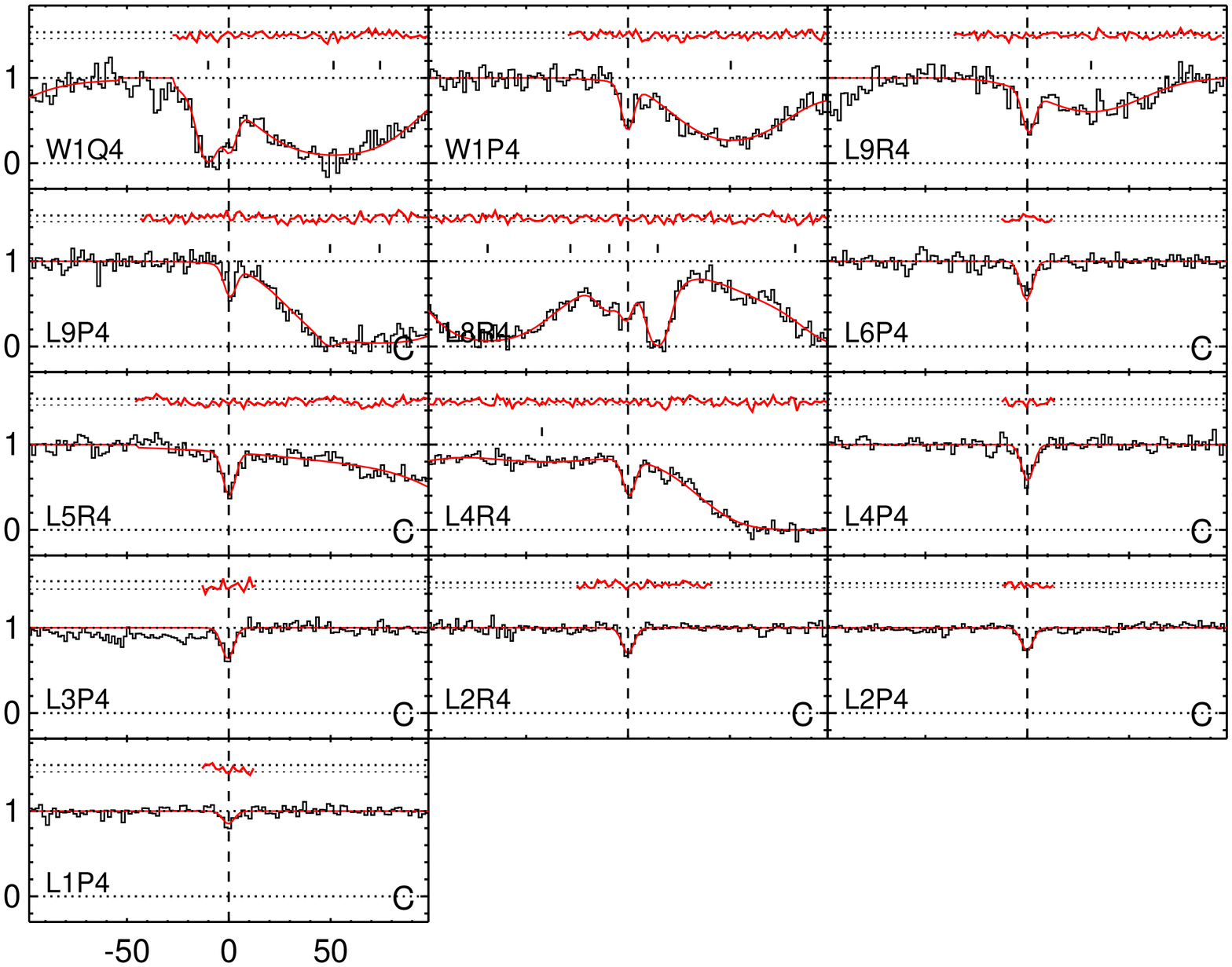}
\caption{The absorption profiles of \h2\ transitions with $J$ = 4 in \sys\ and the best fitted Voigt profile 
to the combined spectrum of all exposures after excluding EXP19. The normalized residual 
(i.e. ([data]-[model])/[error]) for each fit is shown in top of each panel along with the 
$1\sigma$  horizontal line.  The vertical ticks mark the 
positions of fitted contamination.}
\vskip -14.0cm
\begin{picture}(400,400)(0,0)
\put( 170, 44){\large Velocity (\kms)}
\end{picture}
\vskip -14.0cm
\begin{picture}(400,400)(0,0)
\put( -29, 225){\rotatebox{90}{\large Normalized Flux}}
\end{picture}

\label{fig_J4}
\end{figure*}%

\begin{figure*} 
\centering
\vbox{
\includegraphics[width=0.6\hsize,bb=18 18 594 774,clip=,angle=90]{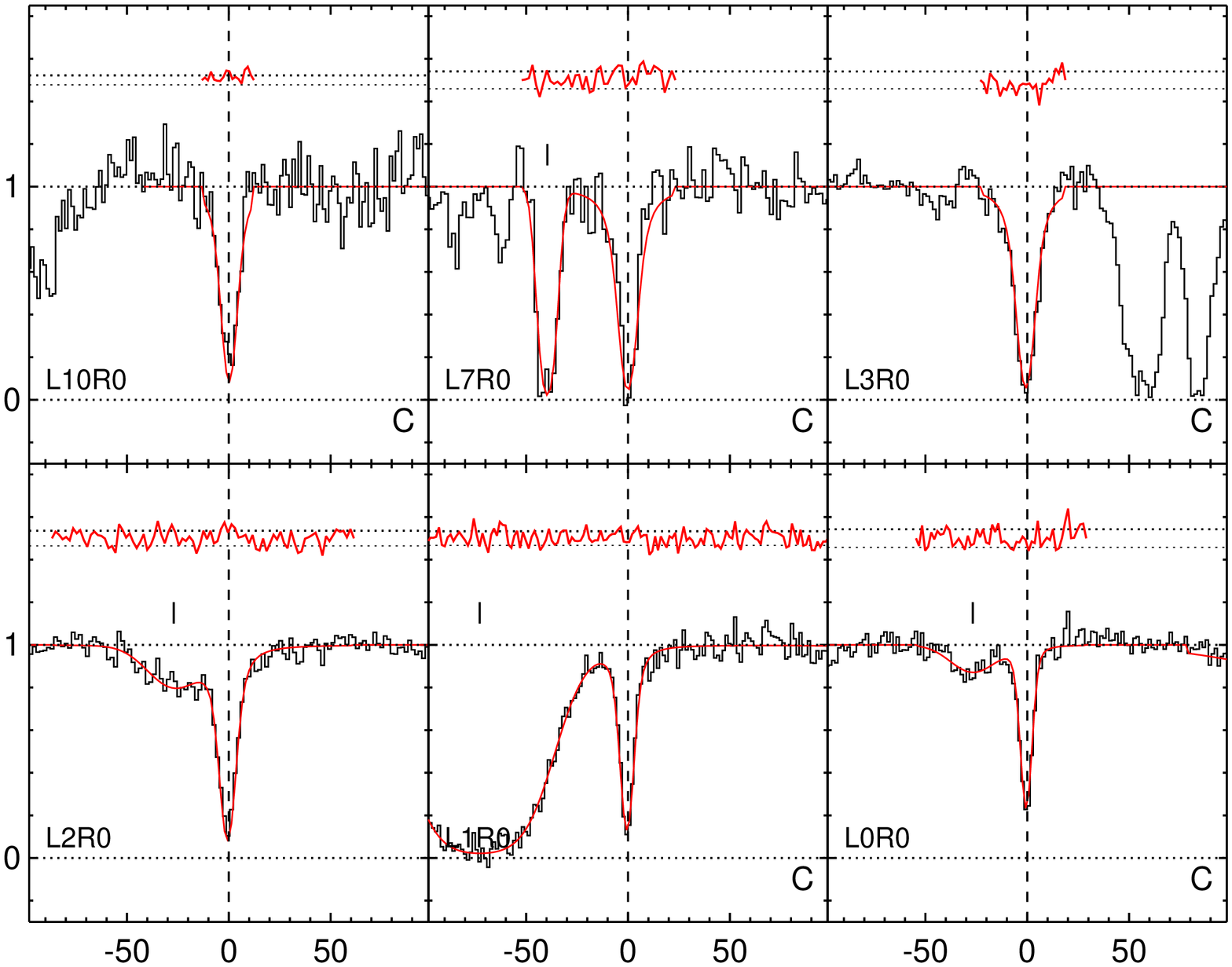}
\includegraphics[width=0.6\hsize,bb=18 18 594 774,clip=,angle=90]{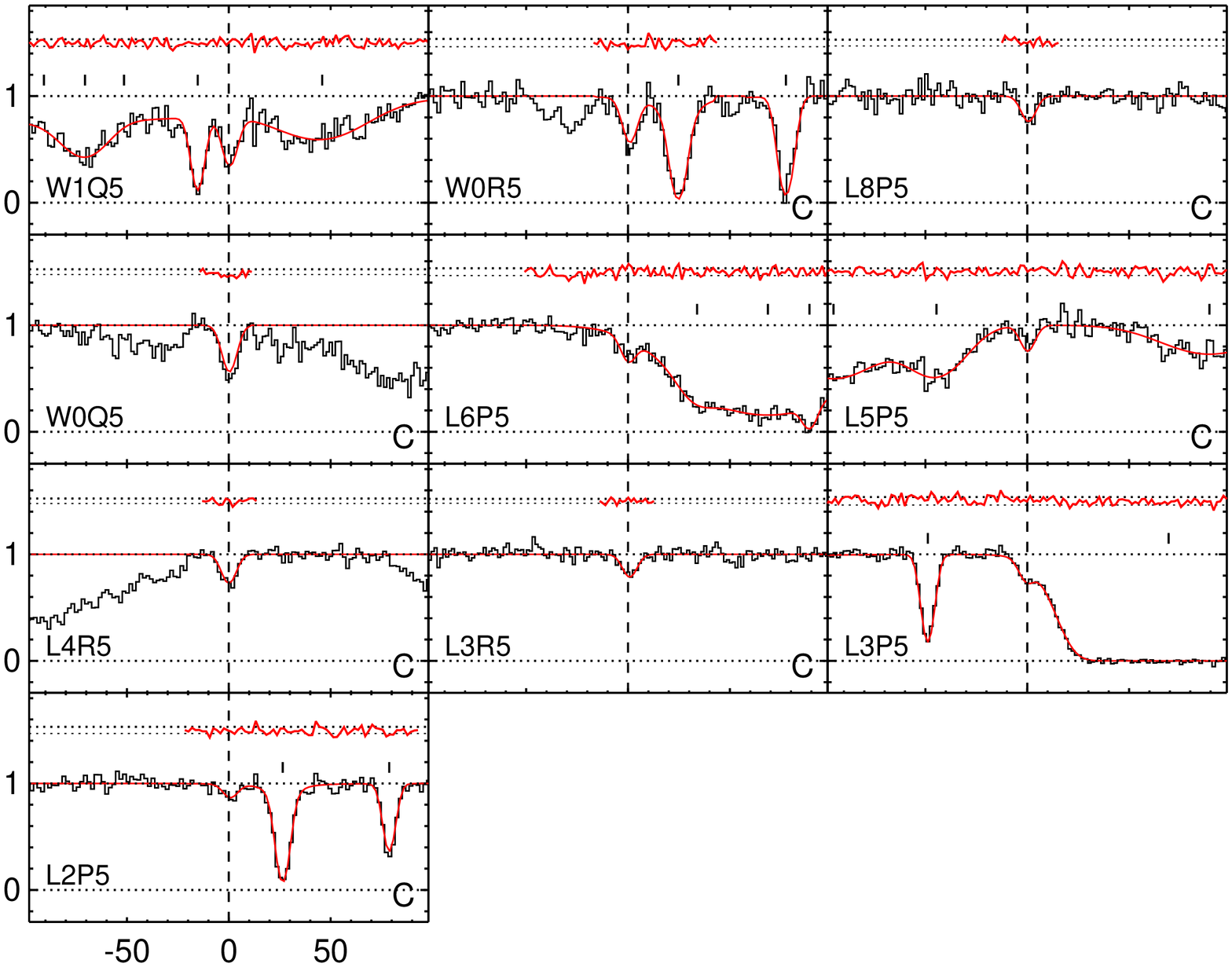}
}
\caption{
The absorption profiles of \h2\ transitions with $J$ = 0 (top) and $J$ = 5 (bottom) in \sys\ and the best fitted Voigt profile 
to the combined spectrum of all exposures after excluding EXP19. The normalized residual 
(i.e. ([data]-[model])/[error]) for each fit is shown in top of each panel along with the 
$1\sigma$  horizontal line.  The vertical ticks mark the 
positions of fitted contamination.}
\label{fig_J0n5}
\vskip -14.0cm
\begin{picture}(400,400)(0,0)
\put( 170, 42){\large Velocity (\kms)}
\end{picture}
\vskip -14.0cm
\begin{picture}(400,400)(0,0)
\put(  10, 160){\rotatebox{90}{\large Normalized Flux}}
\end{picture}
\vskip -14.0cm
\begin{picture}(400,400)(0,0)
\put( 170, 352){\large Velocity (\kms)}
\end{picture}
\vskip -14.0cm
\begin{picture}(400,400)(0,0)
\put(  10, 480){\rotatebox{90}{\large Normalized Flux}}
\end{picture}

\end{figure*}%

\end{document}